\documentclass{aastex}
\usepackage{emulateapj5}
\submitted{Received 2001 October 16; accepted 2001 December 18}
\journalinfo{THE ASTROPHYSICAL JOURNAL, 569, 2002 April 10}
\shortauthors{SAIJO ET AL.}
\shorttitle{COLLAPSE OF A ROTATING SUPERMASSIVE STAR TO A
SUPERMASSIVE BLACK HOLE}
\newlength{\minitwocolumn}
\begin{document}
%
\title
{Collapse of a Rotating Supermassive Star to a Supermassive Black
Hole: Post--Newtonian Simulations}
%
\author
{Motoyuki Saijo \altaffilmark{1},
Thomas W. Baumgarte \altaffilmark{1,2},
Stuart L. Shapiro \altaffilmark{1,3} and
Masaru Shibata \altaffilmark{4}}
%
\affil
{\altaffilmark{1} 
Department of Physics, University of Illinois at Urbana-Champaign,
Urbana, IL 61801-3080}
\affil
{\altaffilmark{2} 
Department of Physics and Astronomy, Bowdoin College,
8800 College Station
Brunswick, ME 04011-8488}
\affil
{\altaffilmark{3} 
Department of Astronomy and NCSA, University of Illinois at
Urbana-Champaign, Urbana, IL 61801}
\affil
{\altaffilmark{4} 
Graduate School of Arts and Sciences, 
University of Tokyo, 
Komaba, Meguro, Tokyo 153-8902, Japan}
%
\begin{abstract}
We study the gravitational collapse of a rotating supermassive star by
means of a (3+1) hydrodynamical simulation in a post-Newtonian
approximation of general relativity.  This problem is particularly
challenging because of the vast dynamical range in space which
must be covered in the course of collapse.  We evolve a uniformly
rotating supermassive star from the onset of radial instability at
$R_{p} / M = 411$, where $R_{p}$ is the  proper polar radius of the star
and $M$ is the total mass-energy,  to the point at which the
post-Newtonian approximation breaks down. We introduce a scale
factor and a ``comoving" coordinate to handle the large variation in
radius during the collapse ($8 \lesssim R_{p}/M_{0} \lesssim 411$,
where $M_{0}$ is the rest mass) and
focus on the central core of the supermassive star.   Since $T/W$, the
ratio of the rotational kinetic energy to the gravitational binding
energy, is nearly proportional to $1/R_{p}$ for an $n=3$ polytropic
star throughout the collapse,  the imploding star may ultimately exceed
the critical value of $T/W$ for dynamical instability to bar-mode
formation.   Analytic estimates suggest that this should occur near
$R_{p}/M \sim 12$, at which point $T/W \sim 0.27$. However, for
stars rotating uniformly at the onset of collapse, we do not find any
unstable growth of bars prior to the termination of our simulation
at $R_{\rm p}/M_{0} \sim 8$.  We do find that the collapse is likely to
form  a supermassive black hole coherently, with almost all of the
matter falling into the hole, leaving very little ejected matter to
form a disk.   In the absence of nonaxisymmetric bar formation,
the collapse of a uniformly rotating supermassive star  does not lead
to appreciable quasi-periodic gravitational wave emission by the time
our integrations terminate.  However, the coherent nature of the
implosion suggests that rotating supermassive star collapse will be a
promising source of gravitational wave bursts.  We also expect
that, following black hole formation, long wavelength quasi-periodic
waves will result from  quasi-normal ringing.   These waves may be
detectable by the Laser Interferometer Space Antenna (LISA). 
\end{abstract}
\keywords{Gravitation --- gravitational waves --- hydrodynamics ---
instabilities ---relativity --- stars: rotation}

\section{Introduction}
\label{sec:intro}
There is increasing evidence that supermassive black holes (SMBHs)
exist at the center of all galaxies, and that they are the sources which
power active galactic nuclei and quasars \citep{Rees98,Macchetto}. 
For example, VLBI observations of the Keplerian disk around an object
in NGC4258 indicate that the central object has a mass $M \sim 3.6
\times 10^{7} M_{\odot}$ and radius less than $\sim 13$pc.  Also,
large numbers of observations are provided by the Hubble space
telescope suggesting that SMBHs exist in galaxies such as M31 ($3
\times 10^{7} M_{\odot}$), M87 ($1 \sim 2 \times 10^{9} M_{\odot}$)
and our own galaxy ($2.5 \times 10^{6} M_{\odot}$) \citep[see for
example,][for a brief overview]{Macchetto}. 

Although evidence of the existence of SMBHs is compelling, the actual 
formation process of these objects is still uncertain \citep{Rees01}.
Several  different scenarios have been proposed, some based on stellar
dynamics, others on gas hydrodynamics, and still others which
combine the processes. In one stellar dynamical scenario, a dense star
system composed of compact stars becomes dynamically unstable to a
collisionless, relativistic radial  mode and undergoes catastrophic
collapse to a SMBH \citep{ZP,ST85a,ST85b,QS87}. In an alternative
scenario,  massive stars build up within a dense cluster, following
collisions and mergers of ordinary stars.  After repeated collisions
and mergers,  supermassive stars (SMSs) are formed, and these become 
unstable to a hydrodynamic, relativistic radial mode 
\citep{Iben,Chandra64a,Chandra64b,Fowler} and eventually collapse to
form SMBHs \citep{Lee,QS90,ccc}.  In still another scenario,  massive halos
of self-interacting dark matter
in the early universe undergo the gravothermal catastrophe (secular core
collapse), followed by catastrophic collapse to SMBHs once their cores become
relativistic \citep{BSI,BS01}.  In a typical gas hydrodynamical
scenario, a contracting primordial gas cloud builds up sufficient 
radiation pressure to inhibit fragmentation, and the gas directly builds
up a SMS \citep*{Sanders,BR,HNR}.  Some mass inevitably will be shed
but most of the matter is trapped during the ensuing gravitational
collapse, forming a SMBH. At present, there is no definitive
observation as yet which confirms or rules out any one of these
scenarios.

Here we focus on the collapse of a SMS.  \citet{BS} investigated the
equilibrium, stability and quasi-static evolution of a SMS with uniform
rotation. During quasi-static evolution, uniform rotation can be
maintained either by internal viscosity or magnetic braking.   They
showed that  the nondimensional ratios $R_{p}/M$, $T/W$ and
$J/M^{2}$ for all critical configurations at the onset of collapse are
universal numbers, independent of the history or mass of the star.  
Here $R_{p}$ is the proper polar radius, $M$ is the gravitational mass
(total mass-energy), $T$ is the rotational kinetic energy, $W$ is the
gravitational binding energy and $J$ is the angular momentum. They
also pointed out the  possibility of bar formation during catastrophic
collapse prior  to BH formation, assuming that the collapse is nearly 
homologous ($T/W \propto 1/R_{p}$).  They provided a crude analytic
argument suggesting that the imploding star should pass the
dynamical bar-mode instability point $T/W \sim 0.27$ at the radius
$R_{p} / M \sim 12$, but were not able to assess the consequences of
this fact.  \citet{NS} later investigated the  quasi-static evolution of
a SMS with differential rotation, assuming negligible viscosity
and magnetic fields.  They showed that in this case, bar formation
prior to the onset of relativistic instability is inevitable.

One of the primary observational missions for space-based detection
of gravitational waves is the investigation of supermassive objects
\citep{Thorne98}.  Since the Laser Interferometer Space Antenna
(LISA) will have long arms ($10^{6} {\rm km}$), the detector will be
most sensitive in the low frequency band ($10^{-4} \sim 10^{-1}$ Hz). 
Potential sources of high signal to noise events in this frequency
range include quasi-periodic waves arising from nonaxisymmetric
bars in collapsing SMSs and the inspiral of binary SMBHs.  In addition,
the nonspherical collapse of rotating SMSs to SMBHs could be a
significant source of burst and quasi-normal ringing radiation. In
this paper we tract the collapse of a SMS by numerical simulation to 
investigate some of these possibilities.

The growth of bars during the catastrophic collapse of stars has been
studied  previously for scenarios leading to core bounce, as in
supernova core collapse.  \citet*{RMR} employed a two-component
equation of state with an effective adiabatic index larger than $4/3$,
which causes bounce.  They concluded that the star forms a bar
in several dynamical timescales after it exceeds the critical
value for  dynamical bar-mode instability ($T/W \gtrsim 0.27$ for
Newtonian  incompressible equilibrium stars).  But since the rotating
core re-expands after the bounce and  $T/W$ falls to a lower value
$\lesssim 0.27$, the nonaxisymmetric instability does not have
sufficient time to enter the nonlinear regime and the bar does not
grow. Therefore, gravitational radiation is not generated effectively in
this core bounce.  \citet{Brown} considered the same problem but chose
a different equation of state to allow the star more time (by a factor
of 100) to reside in the unstable regime $T/W \gtrsim 0.27$.  
Nevertheless, he too found that the dynamical instability was too
weak to produce  an appreciable bar and a large emission of
gravitational waves.  No one has considered the growth of bars during
collapse without a bounce, which is the case for a SMS where the
adiabatic index is close to 4/3.

We take our initial stellar model to be a marginally unstable SMS star
near the critical point, $R_{p}/M \sim 430$.  We treat the gas 
adiabatically, since for sufficiently massive stars neither photon nor
neutrino losses are dynamically significant \citep{LFJMP}.  We take the
adiabatic index to be 4/3, appropriate for a radiation-pressure
dominated SMS, and construct a critical, uniformly rotating  polytrope
with index $n \approx 3$ for our starting point.  Our goal is to
determine the final outcome of the collapse. We want to address the
following questions: Does a SMBH definitely form following the
catastrophic collapse?  Is the collapse coherent or does the central
region collapse first, followed by the gradual accretion of the
envelope?  Does the collapsing  configuration fragment?  Does a disk
form?  Does a rotating bar form during the collapse?  

We use a post-Newtonian (PN) hybrid hydrodynamical code in (3+1)
dimensions to tract SMS collapse. Our adopted hybrid scheme is
relativistically exact for static spherical spacetimes \citep*{SBS98}.
The onset of radial instability occurs when $T/W \ll 0.1$, so our initial
equilibrium spacetime is very nearly spherical. Locating the onset of
radial instability in a SMS requires the presence of nonlinear
gravitation to at least 2PN order \citep{ZN, BS}.  For these reasons,
the nonlinearity captured in our hybrid  scheme, which extends beyond
1PN, is essential to treat this problem.  Of course, it is necessary to
use a fully general relativistic code to follow the final implosion of
the matter into a black hole (BH) and to reliably determine the
gravitational waveforms. (\citet{ST79} followed SMS collapse in full
general relativity for a  nonrotating configuration with a spherical
[1+1] code).  However, a fully relativistic (3+1) code capable of
handling the large dynamic range spanned by SMS collapse is not yet
available. Fortunately, since our initial configuration is nearly
Newtonian, we can use our hybrid scheme to track most of the
implosion up to the point where the formation of a BH is likely.  Our
hybrid scheme is also adequate to address most of  the questions
raised above, at least in a preliminary fashion.

This paper is organized as follows.  In Sec.~\ref{sec:NI} we present the
basic equations of our PN formulation in  ``comoving" coordinates.  We
demonstrate the ability of our code to distinguish stable from
unstable stars in Sec.~\ref{sec:stability}.  We discuss our numerical
results for catastrophic collapse in Sec.~\ref{sec:bar}.   In
Sec.~\ref{sec:Discussion} we summarize our findings.  Throughout this
paper, we use geometrized units ($G=c=1$) and adopt Cartesian
coordinates $(x,y,z)$ with the coordinate time $t$.  Greek and Latin
indices run over $(t, x, y, z)$ and $(x, y, z)$, respectively.

\section{Numerical Method and Key Equations}
\label{sec:NI}
In this section, we briefly derive the (3+1) hybrid PN relativistic
hydrodynamic equations \citep*{SON,SBS98} in ``comoving" coordinates. 
We solve the fully relativistic equations for hydrodynamics, but
neglect some higher-order dynamical PN terms in the Einstein field 
equations.   Note that this approximation gives the exact solution for a
static spherical spacetime.  To track the collapse over the vast 
dynamic range from $\gtrsim 410 M$ down to a few $M$ and to
investigate the central core at late times, we require a suitable
comoving coordinate system.  Such a coordinate choice is possible
because in Newtonian gravity, an $n=3$ spherical polytrope collapses
homologously \citep{GW}.  This special behavior does not strictly hold
in general relativity \citep{ST79}, nor does it apply to a rotating
configuration, but it holds approximately for much of the implosion of
a critical SMS configuration, since it is nearly Newtonian and slowly
rotating.  Therefore, we can construct a ``comoving" frame, subtracting 
the mean ``Hubble" flow from the local velocity, to follow  most of the
collapse with sufficient grid resolution.  

First we construct a PN metric in the ``comoving" frame.  The line
element in the PN formalism is written as \citep*{Chandra65,BDS}
\begin{eqnarray}
ds^{2} &=& g_{\mu\nu} dx^{\mu} dx^{\nu}
\nonumber \\
&=&
(-\alpha^{2} + \beta_{k} \beta^{k}) dt^{2} + 
2 \beta_{k} dx^{k} dt + 
\psi^{4} \delta_{ij} dx^{i} dx^{j}
\label{eqn:1PNmetric}
,
\end{eqnarray}
where $\alpha$, $\beta^{i}$, and $\psi$ is a lapse function, shift
vector, and conformal factor, respectively.  Therefore, as in the
study of BH formation in the Friedman Universe \citep{SS}, we define
the ``comoving" frame by
\begin{equation}
x^{i} = \hat{a} \hat{x}^{i},
\label{eqn:Cbasis}
\end{equation}
where $\hat{x}^{i}$ is the spatial coordinate in the ``comoving" frame,
and $\hat{a}$ is a scale factor which only depends on $t$.  Hereafter,
$\hat{A}$ will represent the quantity $A$ as measured in ``comoving"
coordinates.  Note that we do not change the time coordinate. We write
the PN line element in ``comoving" coordinates as 
\begin{eqnarray}
ds^{2} &=& \hat{g}_{\mu\nu} d\hat{x}^{\mu} d\hat{x}^{\nu}
\nonumber \\
&=&
(-\hat{\alpha}^{2} + \hat{\beta}_{k} \hat{\beta}^{k}) dt^{2} + 
2 \hat{\beta}_{k} d\hat{x}^{k} dt + 
\hat{a}^{2} \hat{\psi}^{4} \delta_{ij} d\hat{x}^{i} d\hat{x}^{j}
\label{eqn:1PNmetricH}
.
\end{eqnarray}
By matching the two line elements (eqs. [\ref{eqn:1PNmetric}] and
[\ref{eqn:1PNmetricH}]) and using equation (\ref{eqn:Cbasis}), we
identify
\begin{eqnarray}
\hat{\alpha} &=& \alpha
\label{eqn:Ralpha}
,\\
\hat{\beta}^{k} &=& 
\frac{\beta^{k}}{\hat{a}} + H \hat{x}^{k}
\label{eqn:Rbeta}
,\\
\hat{\psi} &=& \psi
\label{eqn:Rpsi}
,
\end{eqnarray}
where $H\equiv \dot{\hat{a}}/\hat{a}$.  Note that, among the geometric
quantities, we only need to adjust the shift vector.

For a perfect fluid, the energy-momentum tensor in ``comoving"
coordinates is written as 
\begin{equation}
\hat{T}_{\mu \nu} = 
\rho \left( 1 + \epsilon + \frac{P}{\rho} \right) 
\hat{u}_{\mu} \hat{u}_{\nu} + P \hat{g}_{\mu \nu}
,
\end{equation}
where $\rho$, $\epsilon$, and $P$ are rest mass density, specific
internal energy density, and pressure, respectively.   We adopt a
$\Gamma$-law equation in the form $P=(\Gamma-1)\rho \epsilon$ in
this paper.

In the ``comoving" frame, the continuity equation, energy equation,
and Euler equation with $\Gamma$-law equation of state including
artificial viscosity are  written as 
\begin{eqnarray}
\lefteqn{
\frac{\partial (\hat{a}^{3} \rho_{*})}{\partial t} +
\frac{\partial}{\partial \hat{x}^{i}} ( \hat{a}^{3} \rho_{*} \hat{v}^{i})
= 0
,}
\\
\lefteqn{
\frac{\partial (\hat{a}^{3} e_{*})}{\partial t} +
\frac{\partial}{\partial \hat{x}^{i}} (\hat{a}^{3} e_{*} \hat{v}^{i})
}
\nonumber\\
&&=
- \frac{\hat{a}^{3}}{\Gamma}(\rho \epsilon)^{-1+1/\Gamma} 
P_{\rm vis}
\frac{\partial}{\partial \hat{x}^{i}} ( \alpha \hat{u}^{t} \psi^{6}
\hat{v}^{i} )
,\\
\lefteqn{
\frac{\partial}{\partial t} (\hat{a}^{3} \rho_{*} \tilde{\hat{u}}_{i}) +
\frac{\partial}{\partial \hat{x}^{j}} 
(\hat{a}^{3} \rho_{*} \tilde{\hat{u}}_{i} \hat{v}^{j})
}
\nonumber\\
&&=
- \hat{a}^{3} \alpha \psi^{6} 
\frac{\partial}{\partial \hat{x}^{i}} (P+P_{\rm vis}) -
\hat{a}^{3} \rho_{*} \alpha \tilde{\hat{u}}^{t}
\frac{\partial \alpha}{\partial \hat{x}^{i}} +
\hat{a}^{3} \rho_{*} \tilde{\hat{u}}_{j}
\frac{\partial \hat{\beta}^{j}}{\partial \hat{x}^{i}} 
\nonumber\\
&&
+ 2 \hat{a}^{3} \rho_{*} \frac{1+ \Gamma \epsilon}{\hat{u}^{t}\psi}
[(\alpha \hat{u}^{t})^{2} -1] \frac{\partial \psi}{\partial \hat{x}^{i}}
,
\label{eqn:CEuler}
\end{eqnarray}
where
\begin{eqnarray}
\rho_{*} &=& 
\rho \alpha \hat{u}^{t} \psi^{6}
,\\
\hat{v}^{i} &=& \frac{\hat{u}^{i}}{\hat{u}^{t}}
,\\
e_{*} &=&
( \rho \epsilon)^{1/\Gamma} 
\alpha \hat{u}^{t} \psi^{6}
,\\
\tilde{\hat{u}}^{t} &=& (1+\Gamma\epsilon) \hat{u}^{t}
,\\
\tilde{\hat{u}}_{i} &=& (1+\Gamma\epsilon) \hat{u}_{i}
.
\end{eqnarray}
Following \citet{Shibata99} we include artificial viscosity in our
system as
\begin{equation}
P_{\rm vis} =
\cases{ 
C_{\rm vis} 
\frac{\displaystyle a^{3} e_{*}^{\Gamma}}{\displaystyle (\alpha
\hat{u}^{t}
\psi^{6})^{\Gamma-1}} (\delta \hat{v})^{2},
& for $\delta \hat{v} \leq 0$;
\cr
0, & for $\delta \hat{v} \geq 0$,\cr
}
\end{equation}
with $\delta \hat{v} = 2 \delta \hat{x} \hat{\partial}_{i} \hat{v}^{i}$,
$C_{\rm vis} = 1$, and $\delta \hat{x}$ is the step size of the 
``comoving" grid.  

Gravitational field equations in the PN approximation are derived
from the Hamiltonian constraint,  momentum constraint and the
maximal time-slicing condition \citep*{SON,SBS98}.  The equations in
the comoving frame are written as
\begin{eqnarray}
&&\hat{\triangle} \psi =
- 2 \pi \hat{a}^{2} \psi^{5} \hat{\rho}_{\rm H}
,\\
&&\hat{\triangle} ( \alpha \psi ) =
2 \pi \alpha \hat{a}^{2} \psi^{5} (\hat{\rho}_{\rm H} + 2\hat{S})
,\\
&&\delta_{ij} \hat{\triangle} \hat{\beta^{j}} + \frac{1}{3}
\hat{\partial}_{i} \hat{\partial}_{j} \hat{\beta}^{j} =
16 \pi \alpha \hat{J}_{i}
\label{eqn:Rmomentum}
,
\end{eqnarray}
where $\hat{\rho}_{\rm H} = n_{\mu} n_{\nu} \hat{T}^{\mu \nu}$,
$\hat{J}_{i} = - n_{\mu} \hat{h}_{i \nu} \hat{T}^{\mu \nu}$, $\hat{S} =
\hat{h}^{i}_{\mu} \hat{h}_{i\nu} \hat{T}^{\mu \nu}$,
$n_{\mu} = (-\alpha,0,0,0)$, $\hat{h}_{\mu \nu} = \hat{g}_{\mu \nu} +
n_{\mu} n_{\nu}$, and $\hat{\triangle}$ is the flat Laplacian measured
in the ``comoving" frame.

We use the asymptotic fall-off behavior for metric quantities at
large radius in order to set an appropriate boundary condition at the
grid edge \citep{SBS98}. Only the rescaled shift vector
$\hat{\beta}^{i}$ requires  special consideration. 
Although $\beta^{i}$ has a simple fall-off at the large
radii, falling like $\sim O(r^{-2})$, $\hat{\beta}^{i}$ diverges
like
$\sim H \hat{x}^{i}$.  In order to eliminate the divergent behavior of
$\hat{\beta^{i}}$, we introduce a new vector $\hat{W}^{i}$ as
\begin{equation}
\hat{W}^{i} \equiv \hat{\beta}^{i} - H \hat{x}^{i}
.
\end{equation}
In terms of $\hat{W}^{i}$, the momentum constraint equation (eq.
[\ref{eqn:Rmomentum}]) can be rewritten as
\begin{equation}
\delta_{ij} \hat{\triangle} \hat{W}^{j} + \frac{1}{3}
\hat{\partial}_{i} \hat{\partial}_{j} \hat{W}^{j} =
16 \pi \alpha \hat{J}_{i}
.
\end{equation}
From a computational point of view, it is useful to split this
momentum equation into one vector Poisson equation and one
scalar Poisson equation using the technique of \citet{Shibata97}: 
\begin{equation}
\delta_{ij} \hat{W}^{j} = 4 \hat{B}_{i} - 
\frac{1}{2} [\hat{\partial}_{i} \hat{\chi} + \hat{\partial}_{i}
(\hat{B}_{k} \hat{x}^{k})]
,
\end{equation}
where $\hat{B}_{i}$ and $\hat{\chi}$ satisfy
\begin{eqnarray}
\hat{\triangle} \hat{B}_{i} &=& 4 \pi \alpha \hat{J}_{i}
,\\
\hat{\triangle} \hat{\chi} &=& 
- 4 \pi \alpha \hat{J}{_{i}} \hat{x}^{i}
.
\end{eqnarray}
We also need to rewrite the Euler equation [eq. (\ref{eqn:CEuler})] using
$\hat{W}^{i}$ as
\begin{eqnarray}
\lefteqn{\frac{\partial}{\partial t}
( \hat{a}^{2} \rho_{*} \tilde{\hat{u}}_{i}) +
\frac{\partial}{\partial \hat{x}^{j}} 
( \hat{a}^{2} \rho_{*} \tilde{\hat{u}}_{i} \hat{v}^{j})
}
\nonumber\\
&=&
- \hat{a}^{2} \alpha \psi^{6} 
\frac{\partial}{\partial \hat{x}^{i}} (P+P_{\rm vis})-
\hat{a}^{2} \rho_{*} \tilde{\hat{u}}^{t} 
\frac{\partial \alpha}{\partial \hat{x}^{i}} +
\hat{a}^{2} \rho_{*} \tilde{\hat{u}}_{j}
\frac{\partial \hat{W}^{j}}{\partial \hat{x}^{i}} 
\nonumber \\
&&+
2 \hat{a}^{2} \rho_{*} \frac{1+ \Gamma \epsilon}{\hat{u}^{t}\psi}
[(\alpha \hat{u}^{t})^{2}-1] \frac{\partial \psi}{\partial \hat{x}^{i}},
\end{eqnarray}
where
\begin{eqnarray}
\hat{v}^{i} &\equiv& \frac{\hat{u}^{i}}{\hat{u}^{t}}
=
-(\hat{W}^{i} + H \hat{x}^{i} ) + 
\frac{\delta^{ij} \hat{u}_{j}}{\hat{a}^{2} \psi^{4} \hat{u}^{t}}
.
\end{eqnarray}
The fluid velocities in the two frames are related through equation
(\ref{eqn:Cbasis}) by
\begin{equation}
v^{i} = \hat{a} ( \hat{v}^{i} + H \hat{x}^{i} )
.
\end{equation}

Finally, we must decide how to choose $H$.  As long as homology holds
during the collapse, we can always subtract the ``Hubble" flow by
choosing the appropriate $H$ ($\hat{v}^{r}=0$).  This is why the star
does not contract during the collapse in the ``comoving" frame.  Since
rotating collapse does not strictly hold homology due to the effects of
rotation and PN gravity, success in exploring the late collapse depends
on the location where we choose to subtract  the ``Hubble" flow. 
Though an
$n=3$ polytropic star has a large envelope, most of the mass is
located in the central core.  Therefore we set $H$ around the mean
radius according to 
\begin{equation}
H =
\left.
\left(
- \hat{W}^{x} + \frac{\delta^{xj} \hat{u}_{j}}{\hat{a}^{2} \psi^{4}
\hat{u}^{t}}
\right) / \hat{x}^{x}
\right|_{\hat{r} \sim \hat{r}_{m}}
,
\end{equation}
where the mean radius is defined as
\begin{equation}
\hat{r}_{m}=
\sqrt{
\frac{\int \rho_{*} \hat{x}^{2} d^{3}\hat{x}}{\int \rho_{*} d^{3}\hat{x}}
}.
\end{equation}
We experiment by using different radii in the neighborhood of
$r_{m}$ for measuring $H$ and choose the one which yields the most
accurate evolution. Note that for homologous Newtonian collapse of an 
$n=3$ spherical polytrope, $H$ defined above is independent of the
radius at which it is evaluated.

We monitor several global quantities of the system during the collapse.
The gravitational mass $M$, 
rest mass $M_{0}$, proper mass $M_{p}$,  angular
momentum $J$, rotational kinetic energy $T$, and gravitational
binding energy $W$ of the rotating star are defined as
\begin{eqnarray}
M &=& - \frac{1}{2 \pi} \int_{r \rightarrow \infty} \nabla^{i} \psi
dS_{i}
\nonumber \\
&=&
\int \hat{a}^{3} \biggl[ ( \rho + \rho \epsilon + P ) 
(\alpha \hat{u}^{t})^{2} - P
\biggl]
\psi^{5} d^{3}\hat{x}
,\\
M_{0} &\equiv&
\int \rho u^{t} \sqrt{-g} d^{3} x
= \int \hat{a}^{3} \rho_{*} d^{3} \hat{x}
,\\
M_{p} &=&
\int \rho u^{t} ( 1 + \epsilon ) \sqrt{-g} d^{3} x
\nonumber \\
&=& \int \hat{a}^{3} \rho_{*} ( 1 + \epsilon ) d^{3} \hat{x}
,\\
J &=&
\frac{1}{8 \pi} \int_{r \rightarrow \infty} (x K^{i}_{y} - y K^{i}_{x})
\psi^{6} dS_{i}  
\nonumber \\
&=& \int \hat{a}^{3} (\hat{x} \hat{J}_{y} - \hat{y} \hat{J}_{x})
\psi^{6} d^{3}
\hat{x} ,\\
T &=& \frac{1}{2} \int \Omega dJ 
=
\frac{1}{2} \int  \hat{a}^{3} \hat{\Omega} (\hat{x} \hat{J}_{y} - \hat{y}
\hat{J}_{x})
\psi^{6} d^{3}
\hat{x} ,\\
W &=&
| M_{p} + T - M |
,
\end{eqnarray}
where $\Omega$ is the angular velocity of the star,
\begin{equation}
\Omega = \frac{x v^{y} - y v^{x}}{x^{2}+y^{2}}.
\end{equation}
Note that in PN 
gravity, $M$ and $J$ are not strictly conserved, especially when the
field becomes strong and higher order corrections to the gravitational
field become important.  However, these quantities should be well
conserved during those early and intermediate epochs when gravity 
remains weak.  Also $J$ defined above should be strictly conserved
whenever the system is axisymmetric.  Since our system is nearly
axisymmetry, prior to any bar formation, monitoring $J$ conservation
enables us to check the accuracy of our numerical results.  We
also define the quadrupole moment $I_{ij}$ as
\begin{equation}
I_{ij} 
= \int \hat{a}^{5} \rho_{*} \hat{x}^{i} \hat{x}^{j} d^{3} \hat{x}
,
\end{equation}
and the nonaxisymmetric distortion parameter $\eta$ as 
\begin{equation}
\eta = \frac{I_{xx}-I_{yy}}{I_{xx}+I_{yy}}.
\end{equation}

As we choose a polytropic equation of state for our initial data, with
$P = \kappa \rho ^{1+1/n}$, where $n \approx 3$ and $\kappa$ is a 
constant appropriate for thermal radiation pressure of constant
specific entropy \citep[see, e.g.,][eq. {[3]}]{BS}. 
For numerical purposes it is convenient to rescale all quantities  with
respect to
$\kappa$.  Since $\kappa^{n/2}$ has dimensions of length, we introduce
the following nondimensional variables
\begin{equation}
\begin{array}{c c c}
\bar{M} = \kappa^{-n/2} M
, &
\bar{R} = \kappa^{-n/2} R
, &
\bar{J} = \kappa^{-n} J
, \\
\bar{\Omega} = \kappa^{n/2} \Omega
, &
\bar{\rho} = \kappa^{n} \rho
.&
\end{array}
\end{equation}
Therefore our numerical results apply to arbitrary mass, where we
simply need to adjust $\kappa$ according to 
\begin{equation}
\left( \frac{M}{10^{6} M_{\odot}} \right) = 
\left( \frac{\kappa^{n/2}}{1.48 \times 10^{6} \rm{km}} \right) \bar{M}
,
\end{equation}
to apply our result to arbitrary mass.  Henceforth, we adopt
nondimensional quantities, but omit the bars for convenience.
Equivalently, we set $\kappa = 1$.

\section{Stability Analysis}
\label{sec:stability}

Before simulating rotating SMS collapse in our ``comoving" scheme, we
first assess the stability and accuracy of our (3+1) PN hybrid
code.   For this purpose, we study the stability of stars along an
equilibrium sequence of fixed entropy (fixed $\kappa$) using our
dynamical code and compare the results with the exact turning point
criterion. We recall that the onset of radial instability occurs at the
turning point in a plot of $M$ vs. $\rho_c$ along the sequence.

%
%
\subsection{Spherical Symmetric Case}
\label{subsec:spherical}

In general relativity and PN gravitation, all nonrotating spherical
polytropes with index $n=3$ are radially unstable to collapse and have
no turning points along their equilibrium sequences.  To test the ability
of our code to distinguish stable from unstable spherical stars, we
must study polytropes with a slightly smaller $n$; we set $n=2.96$ for
this purpose, for which the equilibrium structure remains close to
that of a radiation-pressure supported SMS.  We perturb the initial
equilibrium at $t=0$ by slightly decreasing the pressure and following
the subsequent evolution.  More specifically, we  decrease the pressure
constant $\kappa$ according to $\kappa \rightarrow 0.99 \kappa$.   We
use  modest grid sizes (typically $99 \times 99 \times 50$)
\footnote{Our code has equatorial plane symmetry.} for this test, 
covering  the star with 81 grid points across its diameter.  

\begin{center}
\begin{minipage}{7.0cm}
\epsfxsize 7.0cm
\epsffile{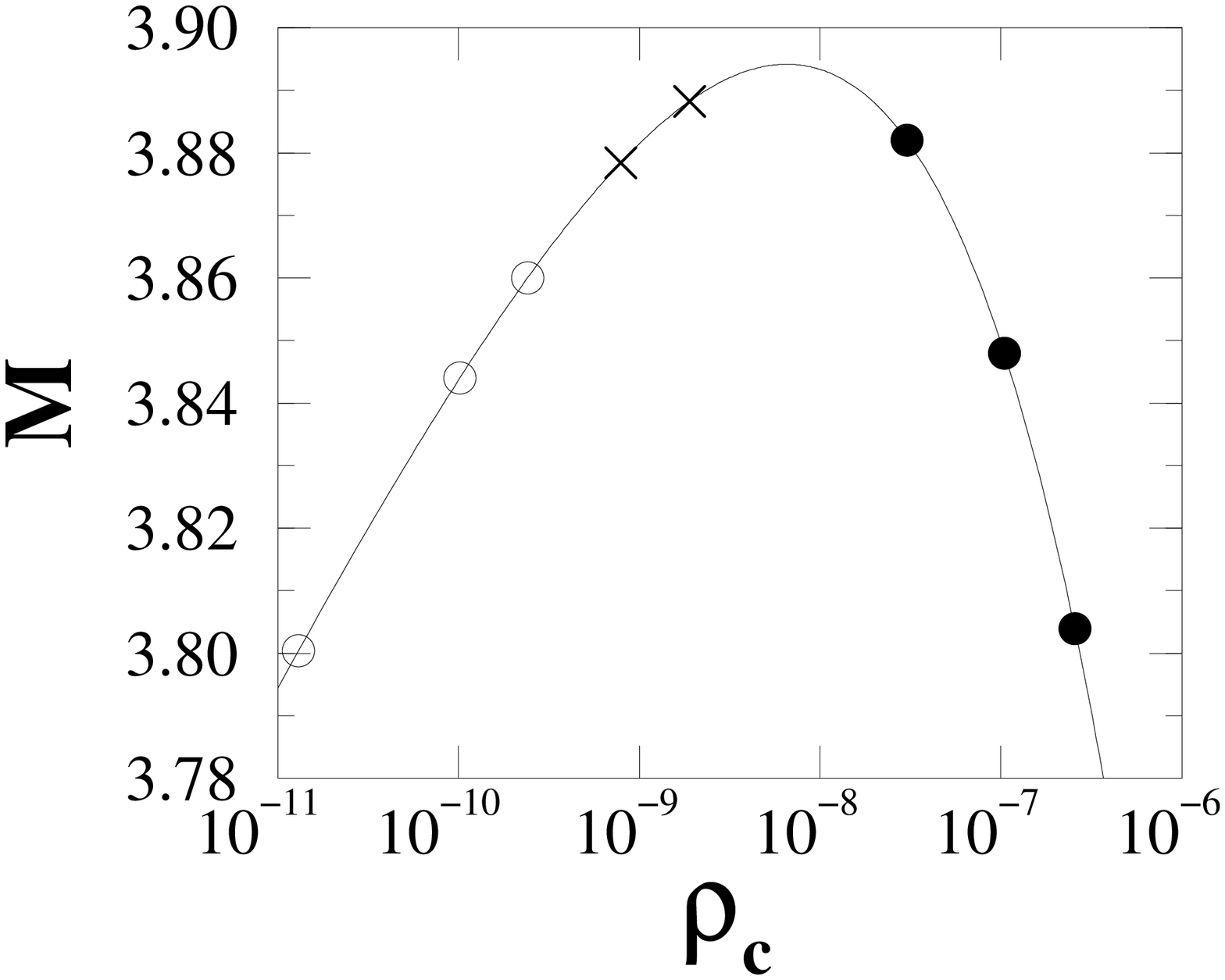}
\end{minipage}
\end{center}
\figcaption[f1.eps]{
Probing the dynamical stability of a spherical SMS with $n=2.96$. 
Here, $\rho_{\rm c}$ is the central density of the equilibrium
spherical star.  Filled circles and crosses represent unstable stars,
while open circles represent stable stars according to our dynamical
calculation.  A cross indicates that the star is actually stable
analytically according to the turning point criterion.   The radii of the
8  marked stars are $R/M = 150$, $200$, $300$, $750$, $1007$,
$1499$, $2001$, $4003$, where the sequence starts at the right side
of the figure at the  highest central densities.  With the adopted grid
resolution, our code can distinguish stable from unstable stars to
within 1\%  of the maximum gravitational mass.
\label{fig:seq}}
\vskip 12pt

Figure \ref{fig:seq} is a summary of our stability code calibration for a
spherical SMS. We find that with the above grid resolution, our code
can distinguish stable from unstable stars to within 1\% of the
maximum gravitational mass. Figure \ref{fig:sstev} compares the
dynamical behavior of stable and unstable stars.  Note that we use the
unit of time as $t_{D} \equiv \sqrt{R_{\rm e}^{3}/M}$ where $R_{\rm
e}$ is the initial equatorial radius ($R_{\rm e}=R$ for a spherical
case).  For dynamically unstable stars the central density increases
monotonically, while for stable stars it oscillates.

%
%
\subsection{Rotating Case}
\label{subsec:rotating}

To assess the ability of our code to distinguish stable from unstable
stars with rotation, we consider an equilibrium sequence of uniformly 
rotating stars of fixed angular momentum $J$ ($J/M^{2} = 0.644$ at 
the turning point). While 
the turning point criterion strictly identifies the onset of secular 
instability \citep*{FIS}, 
\begin{center}
\begin{minipage}{7.0cm}
\epsfxsize 7.0cm
\epsffile{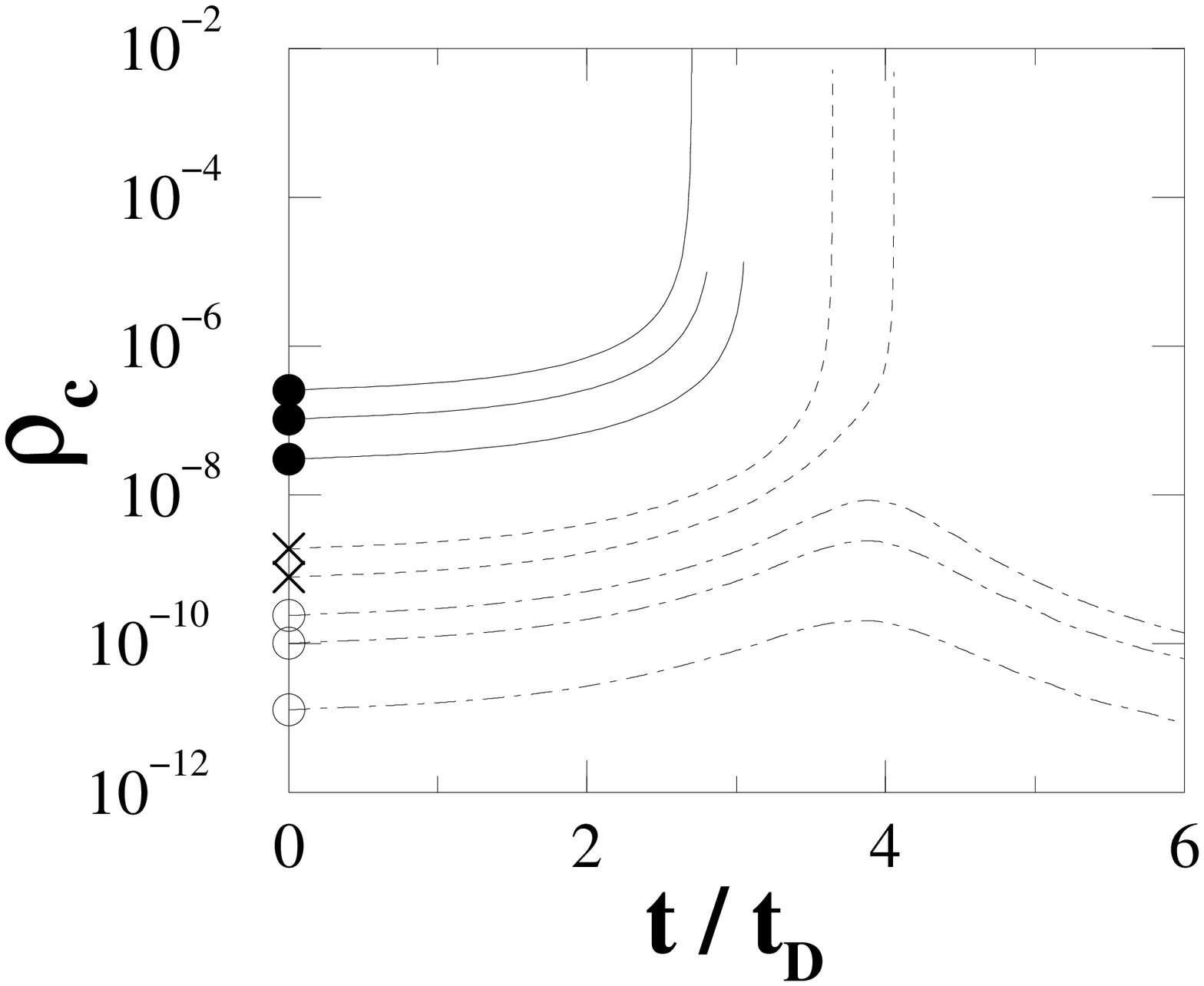}
\end{minipage}
\end{center}
\figcaption[f2.eps]{
Evolution of the central densities of the stars plotted in Fig. 
\ref{fig:seq}.  Here $t_{\rm D}=R_{\rm e}\sqrt{R_{\rm e}/M}$,
where $R_{\rm e}$ is the initial equatorial radius.  Curves are drawn
for stars which are unstable both numerically  and according to the
turning point criterion (solid), unstable numerically but stable
according to the turning point criterion (dashed), and stable both
numerically and according to the turning point criterion (dash-dotted).
\label{fig:sstev}}
\vskip 12pt
the point of onset of dynamical instability 
nearly coincides with the secular instability point \citep*{SBS00}.  
We adopt the same
polytropic index and grid resolution as in the spherical simulations
reported above. We decrease the initial pressure as in the
spherical case ($\kappa \rightarrow 0.99 \kappa$).

\begin{center}
\begin{minipage}{7.0cm}
\epsfxsize 7.0cm
\epsffile{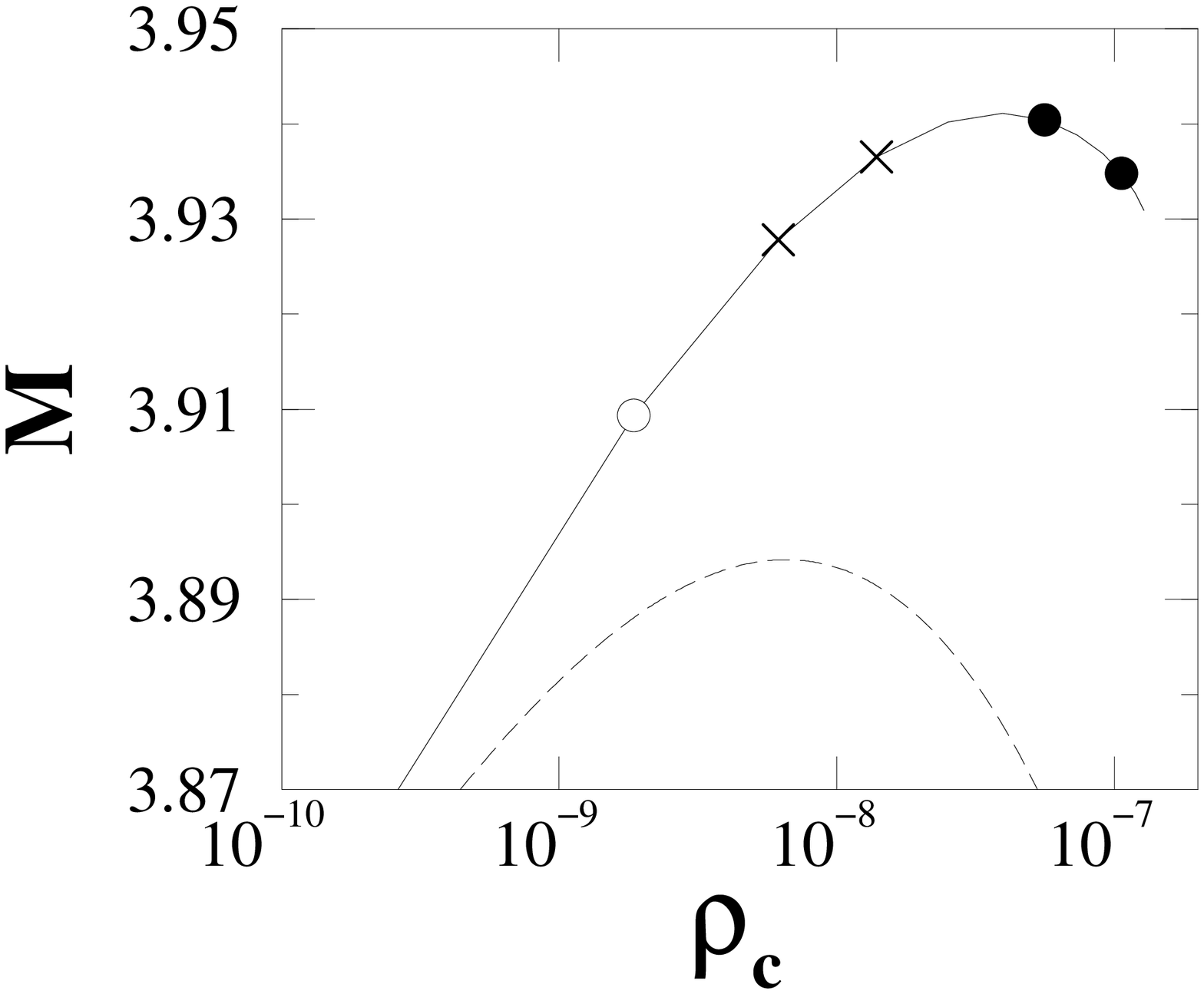}
\end{minipage}
\end{center}
\figcaption[f3.eps]{
Probing the dynamical stability of a rotating SMS with $n=2.96$,
$J=10$. 
Circles and crosses have the same meanings as in Fig. \ref{fig:seq}. 
The radii of the 5 marked stars are $R/M = 252$, $285$, $419$, $537$,
and $785$, where the sequence starts at the right side of the figure
at the  highest central densities.  Note that the solid line shows a 
constant $J$ sequence with $J=10$, while the dashed line represents 
the spherical equilibrium sequence (Fig. \ref{fig:seq}). With the
adopted grid resolution, our code can distinguish stable from unstable
stars to within 1\% of the maximum gravitational mass.
\label{fig:req}}
\vskip 12pt

Figure \ref{fig:req} summarizes our dynamical stability analysis for
the rotating SMS.  We conclude that with the adopted grid resolution, 
our code can distinguish stable from unstable rotating stars to within
1\% of the maximum gravitational mass. Figure \ref{fig:rstev} shows
the evolution of the central density for stable and unstable rotating
stars.

\begin{center}
\begin{minipage}{7.0cm}
\epsfxsize 7.0cm
\epsffile{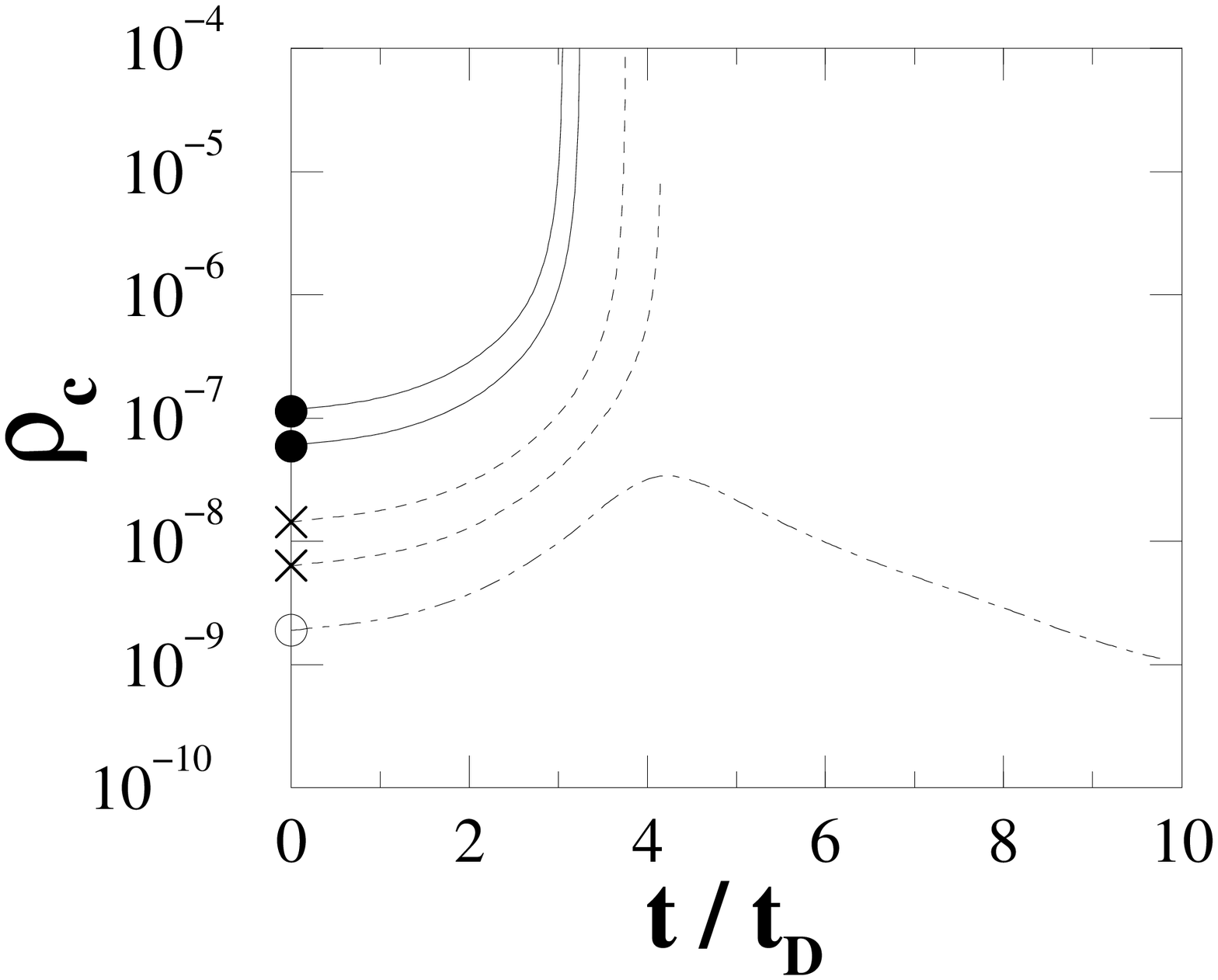}
\end{minipage}
\end{center}
\figcaption[f4.eps]{
Evolution of the central densities of the stars plotted in Fig. \ref{fig:req}.  
Curves are drawn for stars which are unstable both numerically 
and according to the turning point criterion (solid),
unstable numerically but stable according to the turning point criterion
(dashed), and stable both numerically and according to the turning point
criterion (dash-dotted).
\label{fig:rstev}}
\vskip 12pt

\section{Catastrophic Collapse}
\label{sec:bar}
In this section, we simulate catastrophic collapse, focusing on the
possibility of bar formation and on the final fate of the central
object.  First, we examine spherical SMS collapse in PN hybrid
gravitation, to verify numerically that we can recover homology over
the expected  range of radii and that we can resolve the late stages of
the  collapse as long as general relativistic gravity is not too strong. 
Next, we examine  the collapse of a rapidly rotating $n=1$ polytrope to
check that our code can reproduce bar formation when the collapse is
followed by a bounce. Finally, we investigate the collapse of a
uniformly rotating SMS, which is the main result of this paper.  We
discuss the final fate of the SMS collapse, the formation of a SMBH and
the possibility of  circumstellar disk formation and mass loss.

%
%
\subsection{Spherical Collapse}
\label{subsec:scollapse}
We examine spherical SMS collapse to check the range of radii over
which homology holds during the collapse in PN hybrid gravitation.  For
the collapse of an $n=3$  spherical Newtonian star, self-similarity is
strictly preserved \citep{GW}.  In full general relativity the collapse
of an $n=3$ spherical star maintains self-similarity down to the
radius $R/M \lesssim 80$ \citep{ST79}.  Therefore, we expect that
there is some region over which self-similarity is maintained in PN
hybrid gravitation.  Because  $n=3$ spherical stars are marginally
unbound, the system may expand rather than collapse for this
polytropic index when we integrate with finite grid resolution.  To
guarantee collapse, we choose a slightly 
lower index, $n=2.96$ for our
computation.  Since this computation is only to reconfirm homologous
behavior above a minimum stellar radius and to probe how our code
signals the final fate of spherical collapse (a BH), we choose a
modest initial radius $R/M_{0} = 150$.  By performing this spherical
collapse in 3D instead of 1D, this simulation does provide a useful
testbed for our subsequent rotating collapse simulation.  The
parameters characterizing our initial equilibrium star are summarized
in Table \ref{tbl:STInitial}.  We acknowledge that our hybrid PN code
cannot accurately handle BH formation, which is a fully general 
relativistic 
\vskip 12pt
\begin{minipage}{8cm} 
\begin{center}
\tablenum{1}
\label{tbl:STInitial}
\centerline{\sc Table 1}
\centerline{\sc Parameters for the initial}
\centerline{\sc spherical equilibrium SMS.}
\vskip 6pt
\begin{tabular}{c c c}
\hline
\hline
$\bar{\rho}_{c}$ &
$\bar{M}$ &
$R/M$
\\
\hline
$2.56 \times 10^{-7}$ & $3.80$ & $150$
\\
\hline
\hline
\end{tabular}
\end{center}
\vskip 12pt
\end{minipage}
phenomenon.
However, we expect to be able to follow
collapse far enough to ascertain whether or not BH formation is
the likely outcome.

We plot density profiles for the collapsing star at selected times
in Figure \ref{fig:shmg}.  We decrease the initial pressure of the
equilibrium star by 1\% ($\kappa \rightarrow 0.99 \kappa$) to induce
collapse. We see that the collapse is nearly homologous up to a time
$t/t_{\rm D} \sim 2.5$. The radius at that time is $R/M_{0} \sim 80$. 
Note that the central density exceeds 20 times its initial value at this
point. Our result  is consistent with that of \citet{ST79}, who found
that homology was maintained until $R/M \sim 80$.  We do not expect
homologous evolution beyond this point, due to deviations from
Newtonian behavior.  We continue the collapse to $t=2.70 t_{\rm D}$, by
which time the lapse has decreased to $\alpha_{\rm c} \sim 0.2$ and
the mean radius is about $r_{m}/M_{0} \sim 2.0$ (recall that the
horizon radius for a static BH in the adopted isotropic coordinate
system is $r/M = 0.5$). Given the plunge in the lapse, we may  safely
guess that a BH will form after further collapse.

\begin{center}
\begin{minipage}{7.0cm}
\epsfxsize 7.0cm
\epsffile{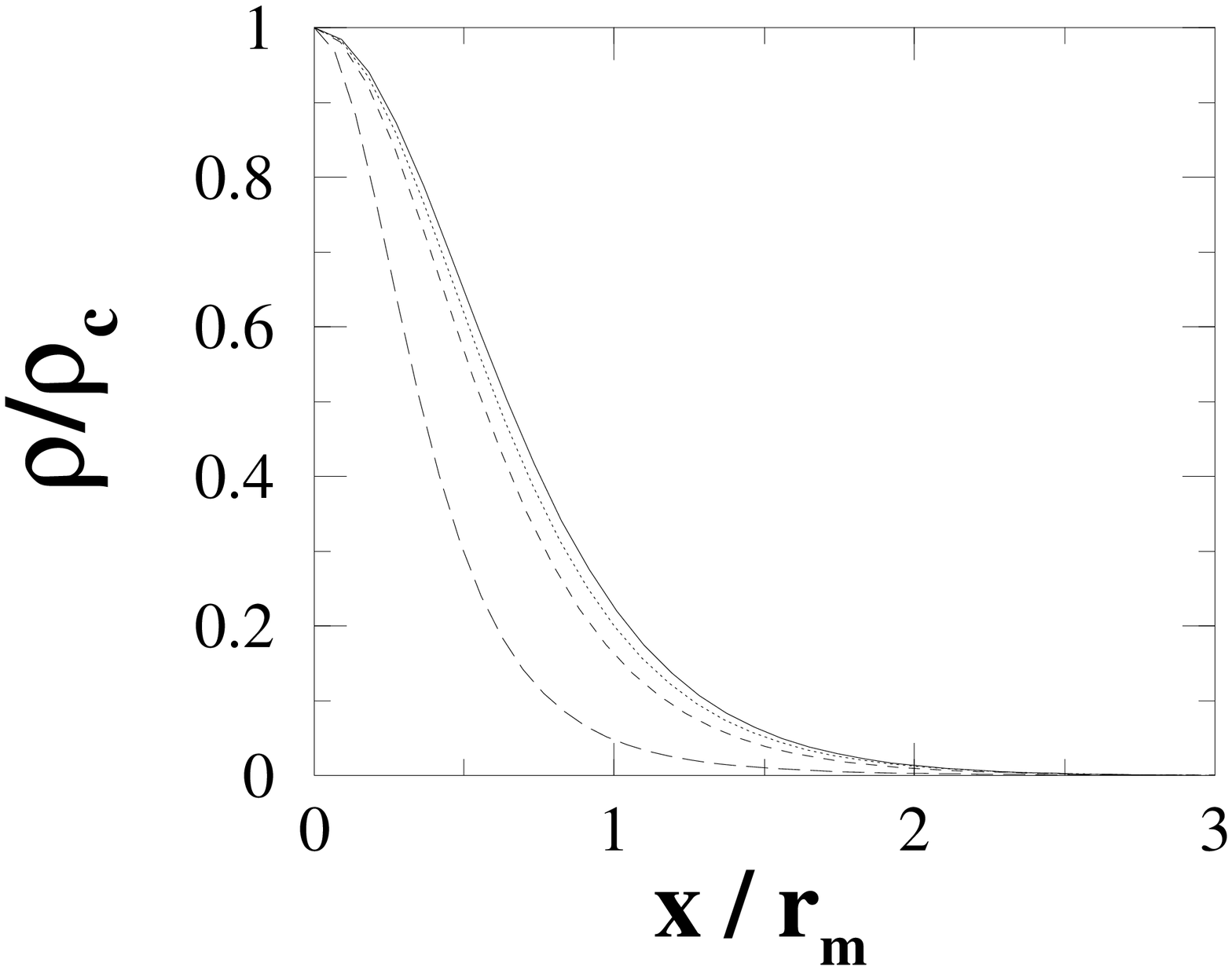}
\end{minipage}
\end{center}
\figcaption[f5.eps]{
Density profiles at selected times during spherical SMS collapse.  
Solid, dotted, dashed, long-dashed lines show profiles at
$t/t_{\rm D} = 0$, $2.04$, $2.52$, $2.68$ and ($\rho_{\rm c} /
\rho_{\rm c}^{(0)}$, $R/M_{0}$) = (1, 146) , (3.05, 131), (16.5, 76.0),
(770, 24.8), respectively.  Note that when a profile overlaps the solid
line, the system is strictly homologous.  Homology is maintained in
this spherical collapse up to the time at which $R/M_{0} \sim 80$.
\label{fig:shmg}}
\vskip 12pt

%
%
\subsection{Core Collapse and Bounce}
One scenario that leads to bar formation is core bounce following the
collapse of a rotating star.  Simulating  this phenomenon tests the
ability of our code to detect a bar-mode instability.  Two previous
numerical studies focus on nonaxisymmetric instabilities  arising in
the core-collapse supernova \citep{RMR,Brown}.  The effective
$\Gamma$ in both cases is larger than $4/3$ at the time of maximum
compression,  which causes the bounce.  They conclude 
that once the
star exceeds the value $T/W \sim 0.27$, bar formation takes place
after several dynamical timescales. We use a similar scenario to test
our code by reproducing

\onecolumn

\begin{minipage}[t]{\minitwocolumn}
this result.  We adopt a stiff $n=1$
($\Gamma = 2$) polytropic model to guarantee a  core bounce following
a modest decrease in initial radius.  Since the change in radius is not
large, we do not need to install a scale factor in this calculation.  As
core bounce will generate a shock between the infalling and outgoing
matter, we must utilize artificial viscosity in our code.

\hspace{3mm}
The initial data for this case is summarized in Table
\ref{tbl:BCInitial}. We choose initial data near the mass-shedding
sequence in order to make the initial value $T/W$ as large as possible. 
We also choose the initial equilibrium star so that the equatorial
radius remains in the PN regime, $R/M \gtrsim 20$. In order to achieve
an appreciable implosion prior to bounce we  drastically deplete the
initial pressure by decreasing $\kappa$ according to
\begin{equation}
\kappa \rightarrow 0.09 \times \kappa.
\label{eqn:Bkappa}
\end{equation} 
We also impose a slight triaxial density perturbation on the
equilibrium star to induce $m=2$ bar formation \citep{RMR}, 
\begin{equation}
\rho = \rho^{\rm (equilibrium)} 
\left( 1 + \delta \frac{x^{2}-y^{2}}{R_{\rm e}^{2}} \right),
\end{equation} 
where $\delta=0.1$.  We employ a grid size ($239 \times 239 \times
120$) which covers the initial equilibrium star with 161 points across
the equatorial diameter.  In fact, the proper equatorial radius varies
between $30 \lesssim R_{\rm e} /M \lesssim 104$ 
during the
\end{minipage}
\hspace{\columnsep}
\begin{minipage}[t]{\minitwocolumn}
collapse, so the resolution of the central region  is always adequate
during the evolution.

\begin{center}
\begin{minipage}[h]{7.0cm}
\epsfxsize 7.0cm
\epsffile{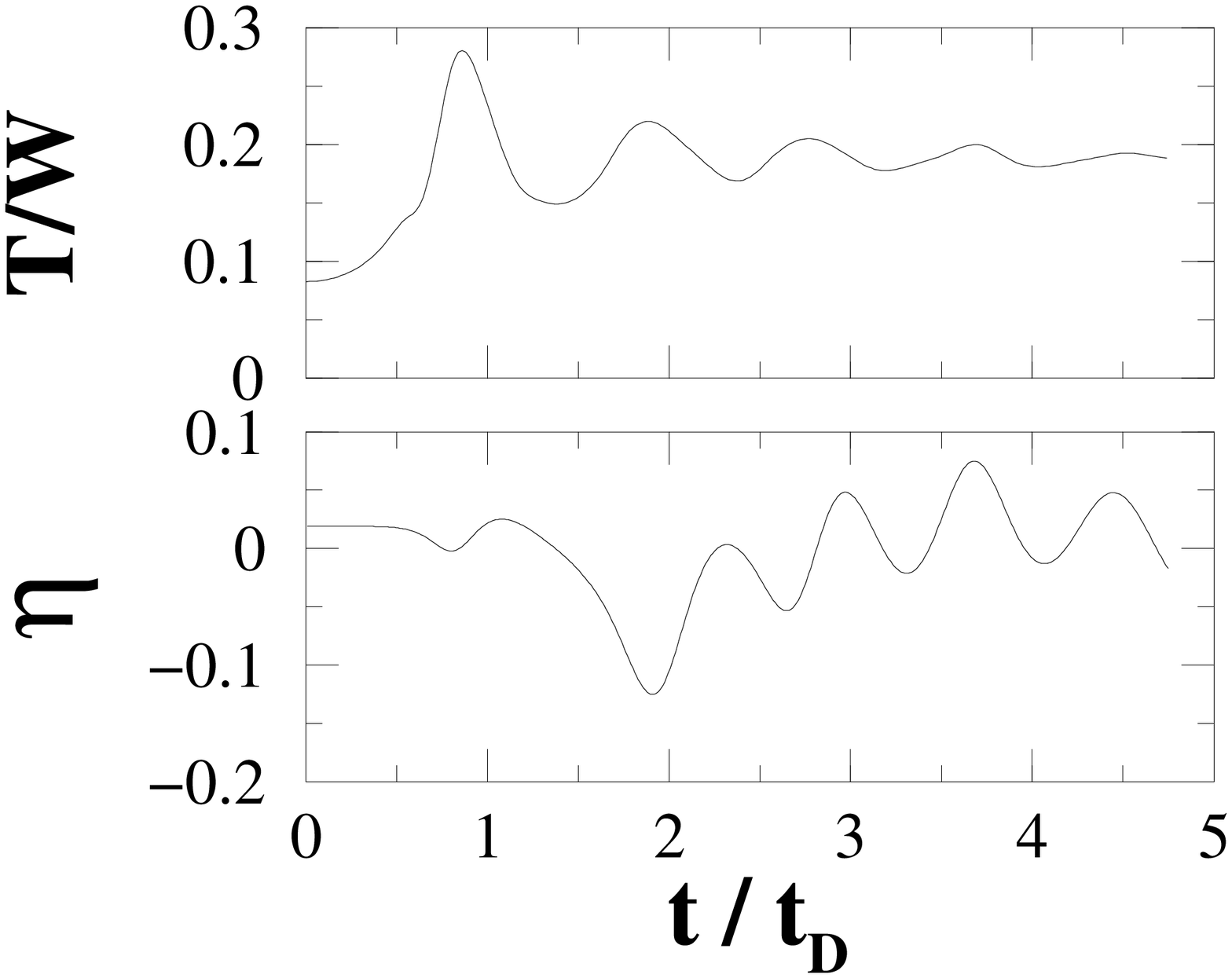}
\end{minipage}
\end{center}
\figcaption[f6.eps]{
Growth of the bar in rotating core collapse and bounce.  The 
quantity $T/W$ exceeds the critical dynamical instability point $\sim 0.27$
during the collapse and settles down below this value to
$T/W \sim 0.19$. The deformation parameter $\eta$ grows 
exponentially during the early evolution following maximum
compression and oscillates about a finite value at
late times.  Note that the behavior of $T/W$ is quite similar to Fig.
3 of \citet*{RMR}.
\label{fig:sk90bar}}
\vskip 12pt
\end{minipage}

\begin{center}
\begin{minipage}{12cm}
\vskip 12pt
\epsfxsize=12cm
\epsffile{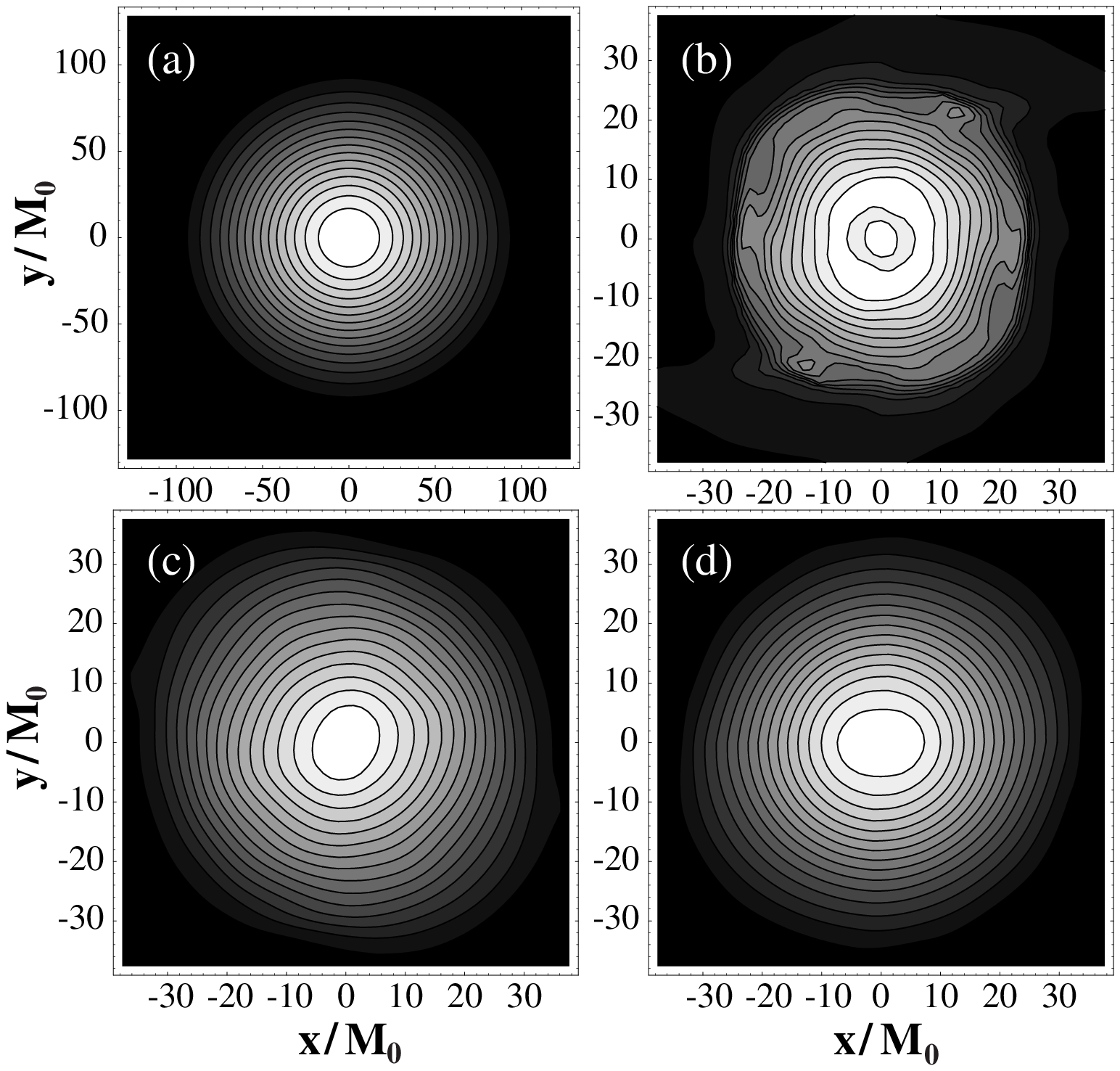}
\end{minipage}
\end{center}
\figcaption[f7.eps]{
Density contours $\rho_{*}$ in the equatorial plane at selected times 
during the collapse and bounce of
a rotating star.
Snapshots are plotted at ($t/t_{\rm D}$, $\rho^{*}_{\rm c}$) $=$ 
(a) ($4.90 \times 10^{-5}$, $5.353 \times 10^{-3}$), 
(b) ($1.41$, $1.410 \times 10^{-1}$), 
(c) ($3.43$, $1.435 \times 10^{-1}$), 
(d) ($4.70$, $1.478 \times 10^{-1}$). 
The contour lines denote densities $\rho^{*}=\rho^{*}_{\rm c} \times
 i/16$ ($i=1, \cdots , 15$).
\label{fig:BCxy}}

\begin{table*}[t]
\begin{center}
\tablenum{2}
\label{tbl:BCInitial}
\centerline{\sc Table 2}
\centerline{\sc Parameters for the initial rotating equilibrium {$n=1$} polytrope.}
\vskip 6pt
\begin{tabular}{c c c c c c c}
\hline
\hline
$\bar{\rho}_{c}$ & $\bar{M}$ & $\bar{J}$ &
$R_{p}/M$ & $J/M^{2}$ & $T/W$ &
$R_{p}/R_{e}$
\\
\hline
$5.00 \times 10^{-3}$ & $1.34 \times 10^{-3}$ &
$4.43 \times 10^{-4}$ & $70.4$ &
$1.91$ & $8.18 \times 10^{-2}$ & $0.675$
\\
\hline
\hline
\end{tabular}
\end{center}
\vskip 12pt
\end{table*}

\begin{table*}[b]
\begin{center}
\tablenum{3}
\label{tbl:RCInitial}
\centerline{\sc Table 3}
\centerline{\sc Parameters for the
initial rotating equilibrium SMS.}
\vskip 6pt
\begin{tabular}{c c c c c c c c}
\hline
\hline
&
$\bar{\rho}_{c}$ & $\bar{M}$ & $\bar{J}$
& $R_{p}/M$ & $J/M^{2}$ & $T/W$ &
$R_{p}/R_{e}$
\\
\hline
Our initial data &$8.00 \times 10^{-9}$ &  $4.57$ & $20.0$ & $411$ &
$0.960$ & $8.85 \times 10^{-3}$ & $0.675$
\\
Critical Value\tablenotemark{a} 
&$7 \times 10^{-9}$ &  $4.57$ & $20.3$ &
$427$ & $0.97$ & $8.99 \times 10^{-3}$ & $0.664$
\\
\hline
\hline
\end{tabular}
\vskip 12pt
\begin{minipage}{13cm}
{${}^{a}${\citet{BS}}}
\end{minipage}
\end{center}
\end{table*}

\newpage

\setlength{\oddsidemargin}{-4mm}
\setlength{\evensidemargin}{-4mm}
\begin{minipage}[t]{\minitwocolumn}
\hspace{3mm}
The bar diagnostics are shown in Figure \ref{fig:sk90bar}.  Our
results are similar to those of \citet{RMR}, especially the behavior of
$T/W$ \citep[Fig. 3 of][]{RMR}. 
Figure \ref{fig:sk90bar} shows the
evolution of the nonaxisymmetric distortion function.  Its growth
signifies bar formation.

\begin{center}
\begin{minipage}{7.0cm}
\epsfxsize 7.0cm
\epsffile{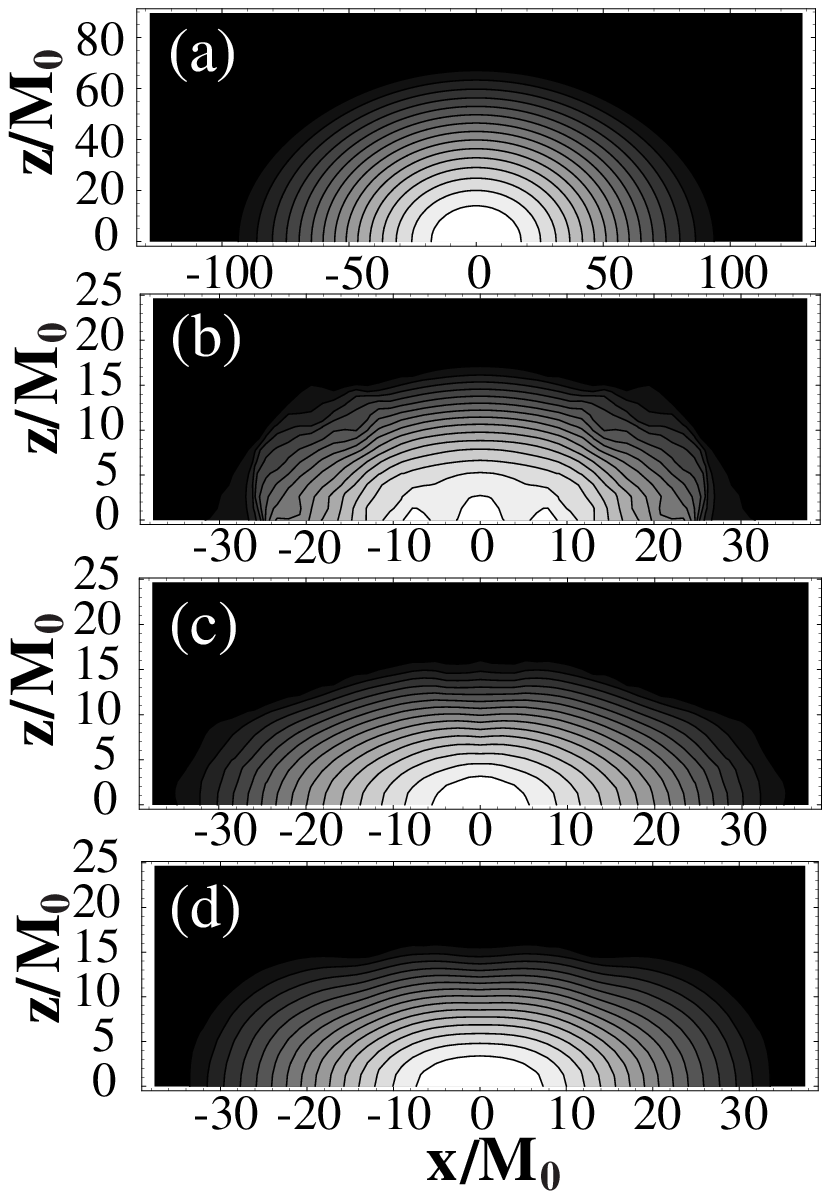}
\end{minipage}
\end{center}
\figcaption[f8.eps]{
Density contours $\rho_{*}$  
in the meridional plane at selected times during the collapse and bounce of
a rotating star.
The times and contour levels are the same as in Fig.
\ref{fig:BCxy}.
\label{fig:BCxz}}
\vskip 12pt
\hspace{3mm}
We plot density contours at selected times in Figures \ref{fig:BCxy} 
(equatorial plane) and \ref{fig:BCxz} (meridional plane).  A modest
spiral arm forms within a dynamical time after the star exceeds
$T/W=0.27$ (Fig. \ref{fig:BCxy} [b]).  After that, the spiral arm is
twisted around the star and settles down to a weakened bar structure
at the central core.  This whole picture
\end{minipage}
\hspace{\columnsep}
\begin{minipage}[t]{\minitwocolumn}
is qualitatively similar to the result reported in \citet{RMR}.

\hspace{3mm}
From the density snapshots and the behavior of distortion function, we
conclude that our code has successfully reproduced (weak) bar 
formation during core collapse and bounce. We thereby confirm that the
code is capable of identifying bars when the physical conditions make
it possible for them to form.

%
%
\subsection{Rotating SMS Collapse}
\label{subsec:rcollapse}
\hspace{3mm}
Consider an overview of SMS evolution \citep{BS}.  Cooling and
contraction of a rotating SMS will ultimately spin it up to the
mass-shedding limit.  After that, the SMS contracts secularly along the
mass-shedding sequence as it cools, slowly losing mass and
maintaining  uniform rotation via viscosity and/or  magnetic braking
\citep{ZN,Shapiro}.   Upon reaching the onset of radial instability, the
star will collapse  catastrophically and form a BH, or a flattened
rotating disk, or some combination thereof. It is this  catastrophic
collapse which we wish to follow with our dynamical code.

\hspace{3mm}
We summarize the 
parameters of our initial uniformly rotating star in Table \ref{tbl:RCInitial}. 
We slightly perturb the initial equilibrium state according to
\begin{eqnarray}
\kappa &\rightarrow& 0.99 \kappa, 
\\
\rho &=& \rho^{\rm (equilibrium)} 
\left( 1 + \delta \frac{x^{2}-y^{2}}{R_{\rm e}^{2}} \right),
\end{eqnarray}
where $\delta=0.1$.  We slightly decrease $\kappa$ in order to deplete
the pressure and initiate the collapse. We install a triaxial density
perturbation to provide the seed for bar formation, if the physical
situation should lead to unstable growth.   We adopt a grid size ($239
\times 239 \times 120$), so 
that the star is initially covered by 161
points across the equatorial diameter. We evolve the rotating SMS up
to the point at which the PN approximation breaks down.

\hspace{3mm}
We show the evolution of the central lapse function in Figure
\ref{fig:qp57valp}.   The figures shows that we can follow the collapse 
from the Newtonian regime where $\alpha_{\rm c} \sim 0.99$ to the
relativistic regime where $\alpha_{\rm c} \sim 0.3$.  The rapid
plummet of $\alpha_{\rm c}$ below $0.3$ at late times indicates that
a BH will form as an immediate consequence of collapse.

\end{minipage}

\newpage
\begin{minipage}[t]{\minitwocolumn}
\hspace{3mm}
We plot the mass and angular momentum during the evolution in Figure 
\ref{fig:qp57vvcons}.  Our hybrid PN scheme conserves $M_{0}$, provided
that the adopted cutoff density in the ambient atmosphere is
negligible
\footnotemark[2]
and that there is no mass loss from the grid.  Our fully relativistic 
expressions for $M$ and $J$, on the other hand, are not necessarily 
conserved in our hybrid PN scheme.  We therefore chose to normalize all 
our length and time units in terms of $M_{0}$.  Toward the end
of our simulation, a small amount ($<4\%$) of the matter
leaves the computational grid, leading to the loss of
rest-mass as shown in Figure \ref{fig:qp57vvcons}.  This figure also 
demonstrates that $M$ and $J$ are conserved to similar levels, indicating
that relativistic effects, which might cause larger deviations, are small.
We show the evolution of the scale factor in Figure
\ref{fig:qp57vscl}.  This plot indicates that physical grid that we
cover shrinks during the collapse  from $x_{\rm max} / M_{0} \sim
900$ ($\hat{a} = 1$) to $x_{\rm max} / M_{0} \sim 50$ ($\hat{a} =
0.06$) where $x_{\rm max}$ is the edge of our numerical grid.

\begin{center}
\begin{minipage}{7.0cm}
\epsfxsize 7.0cm
\epsffile{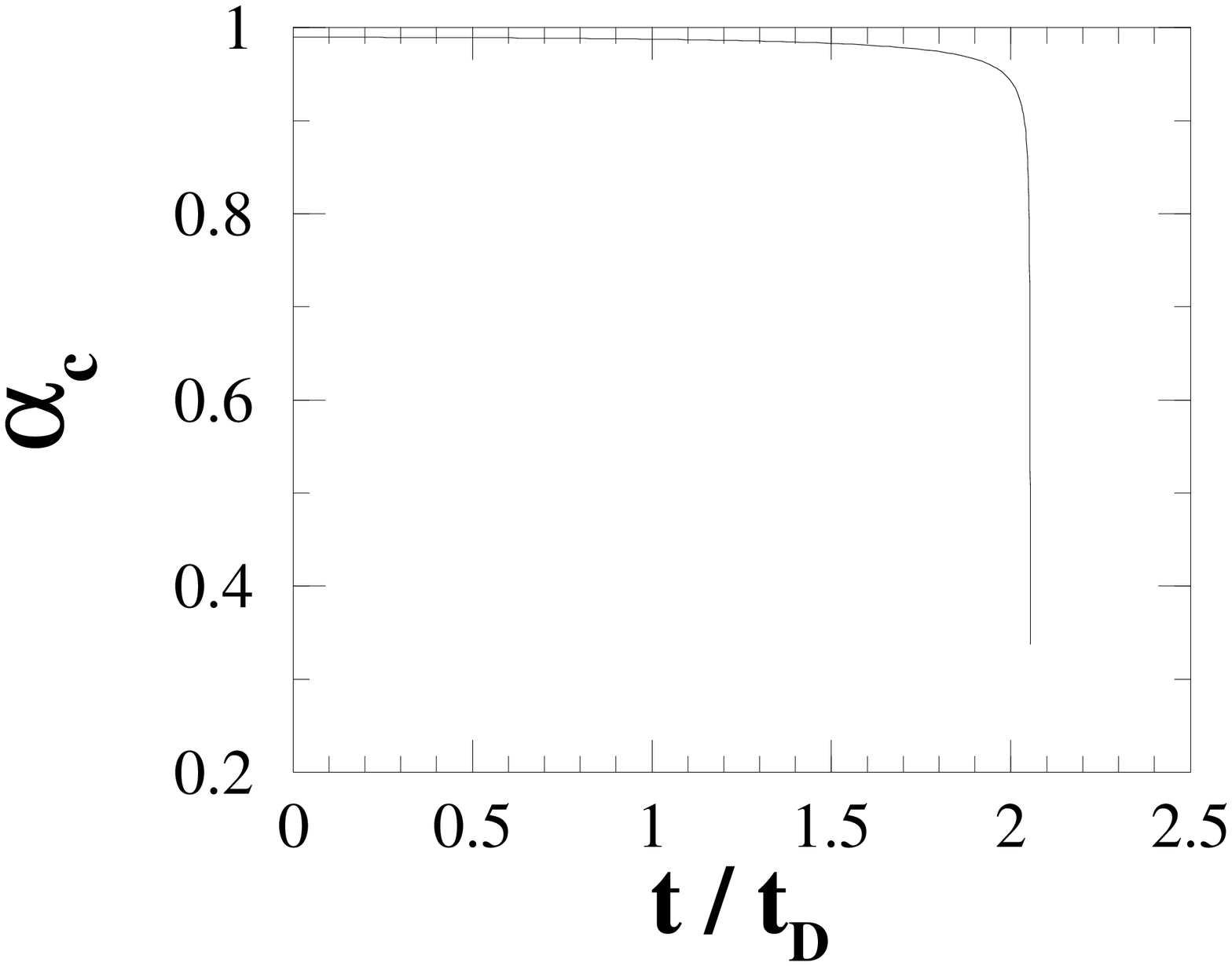}
\end{minipage}
\end{center}
\figcaption[f9.eps]{
Evolution of the central lapse function during rotating SMS collapse.  
\label{fig:qp57valp}}
\vskip 12pt

\begin{center}
\begin{minipage}{7.0cm}
\epsfxsize 7.0cm
\epsffile{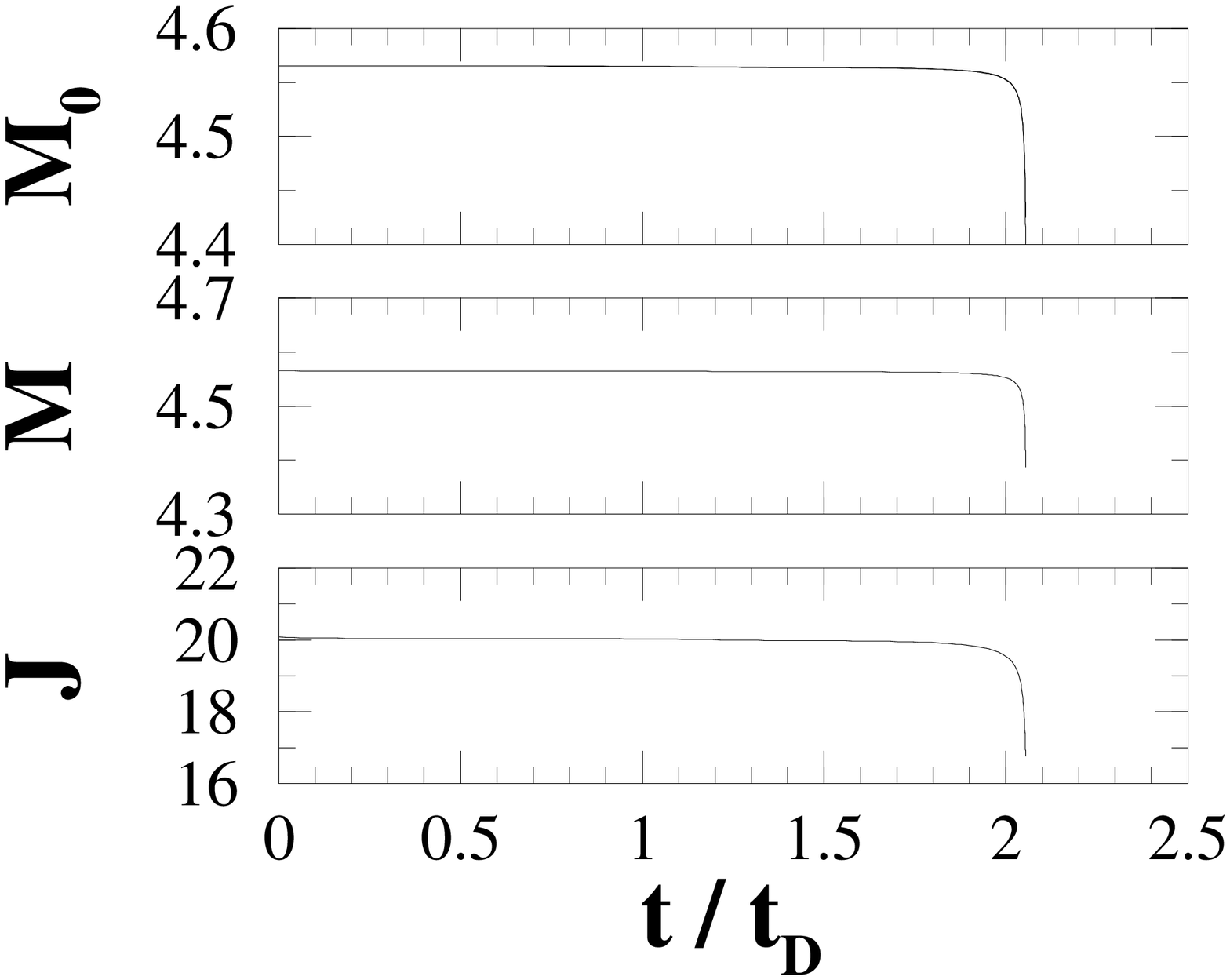}
\end{minipage}
\end{center}
\figcaption[f10.eps]{
Evolution of rest mass $M_{0}$, gravitational mass $M$, and total
angular momentum $J$ during rotating SMS collapse.  Note that 
$M_{0}$ should be strictly
conserved during the evolution {within a numerical error} and it is
indeed conserved within 4\% error (see text).
\label{fig:qp57vvcons}}
\vskip 12pt

\footnotetext{\footnotesize{${}^{2}$In order to regularize the  fluid equations near the surface 
of the star it is convenient to put an atmosphere in the vacuum.  We do 
this by setting a cutoff density below which $\rho^{*}$ may not drop.  
We define the surface of the star to be the place where
$\rho^{*}=\rho^{*}_{\rm s}=10 \rho_{\rm cut}^{*}$, where
$\rho^{*}_{\rm cut}/\rho^{*}_{\rm c} \approx 10^{-9}$.}}

\end{minipage}
\hspace{\columnsep}
\begin{minipage}[t]{\minitwocolumn}
\hspace{3mm}
We monitor the bar-mode diagnostics in Figures \ref{fig:RCbar} and
\ref{fig:TWscale}.  
The amplitude of the deformation function {\it decreases}

\begin{center}
\begin{minipage}{7.0cm}
\epsfxsize 7.0cm
\epsffile{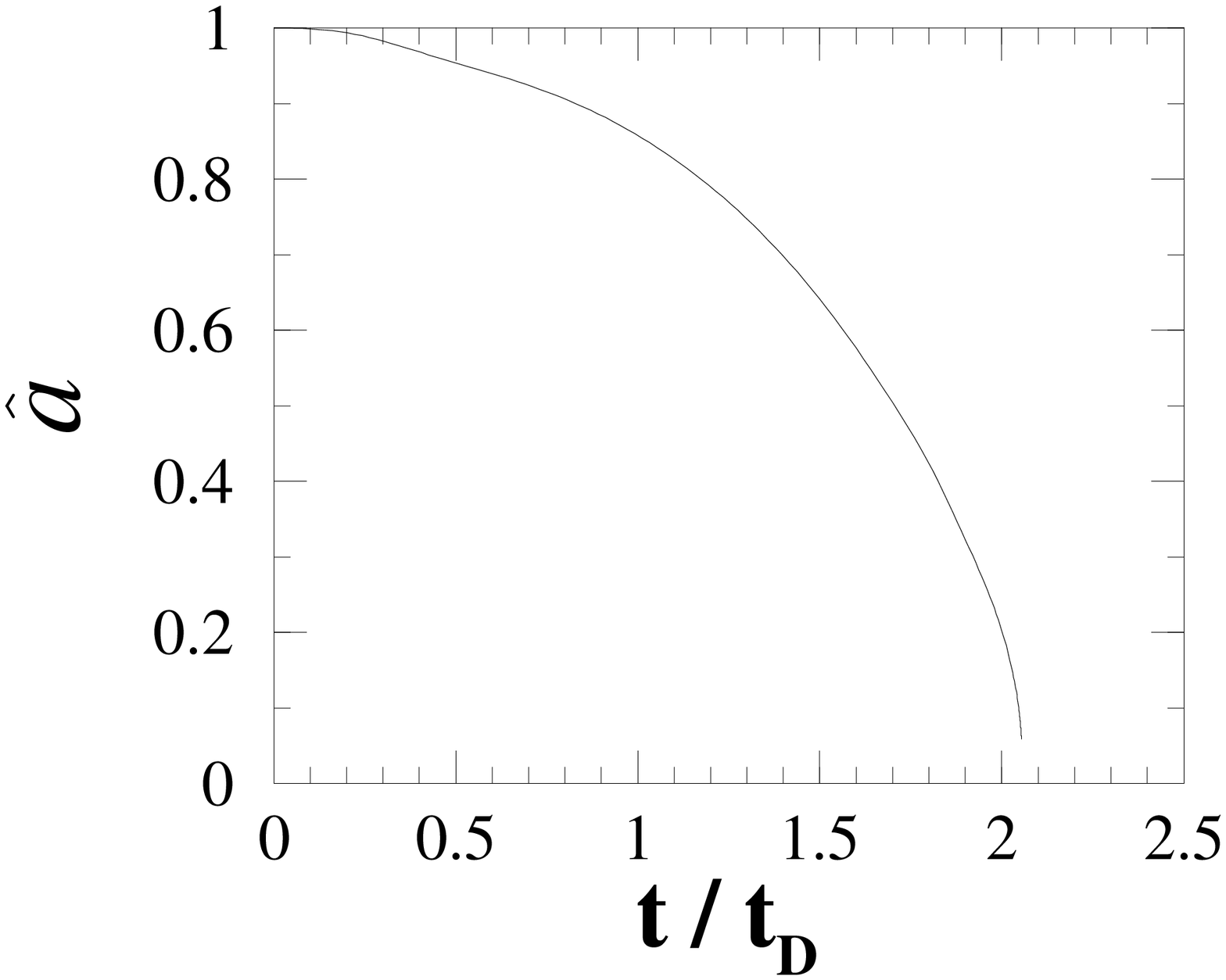}
\end{minipage}
\end{center}
\figcaption[fig11.eps]{
Evolution of the scale factor during rotating SMS
collapse.  The collapse of the scale factor indicates that
the physical size of our grid contracts from
$R/M_{0} \sim 900$ ($\hat{a}=1$) to $R/M_{0} \sim 50$ ($\hat{a}=0.06$).
\label{fig:qp57vscl}}
\vskip 8pt

\begin{center}
\begin{minipage}{7.0cm}
\epsfxsize 7.0cm
\epsffile{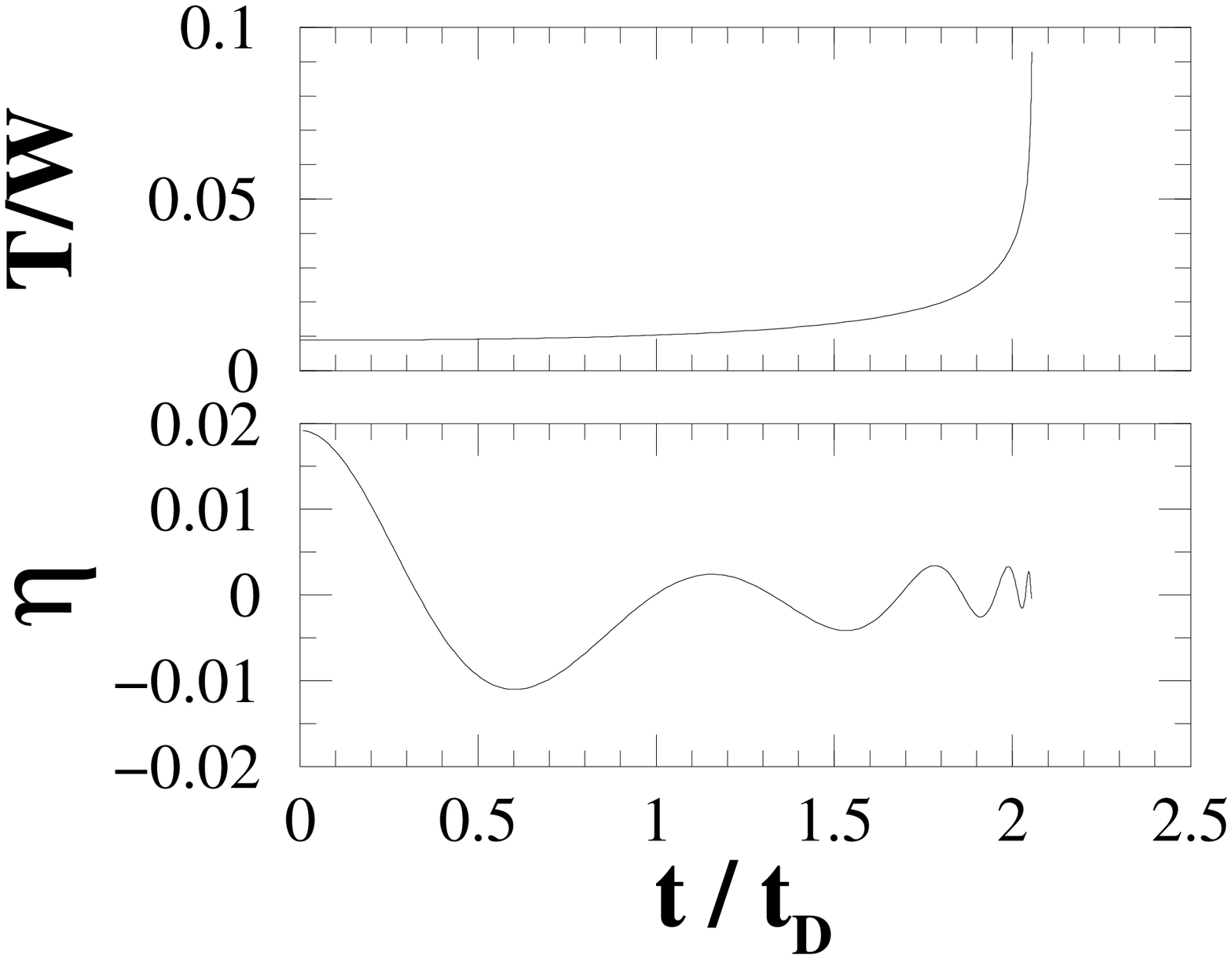}
\end{minipage}
\end{center}
\figcaption[f12.eps]{
Evolution of bar-mode diagnostics during rotating SMS
collapse.  The deformation parameter $\eta$ does not grow 
exponentially even in the final stage.
\label{fig:RCbar}}
\vskip 8pt

\begin{center}
\begin{minipage}{7.0cm}
\epsfxsize 7.0cm
\epsffile{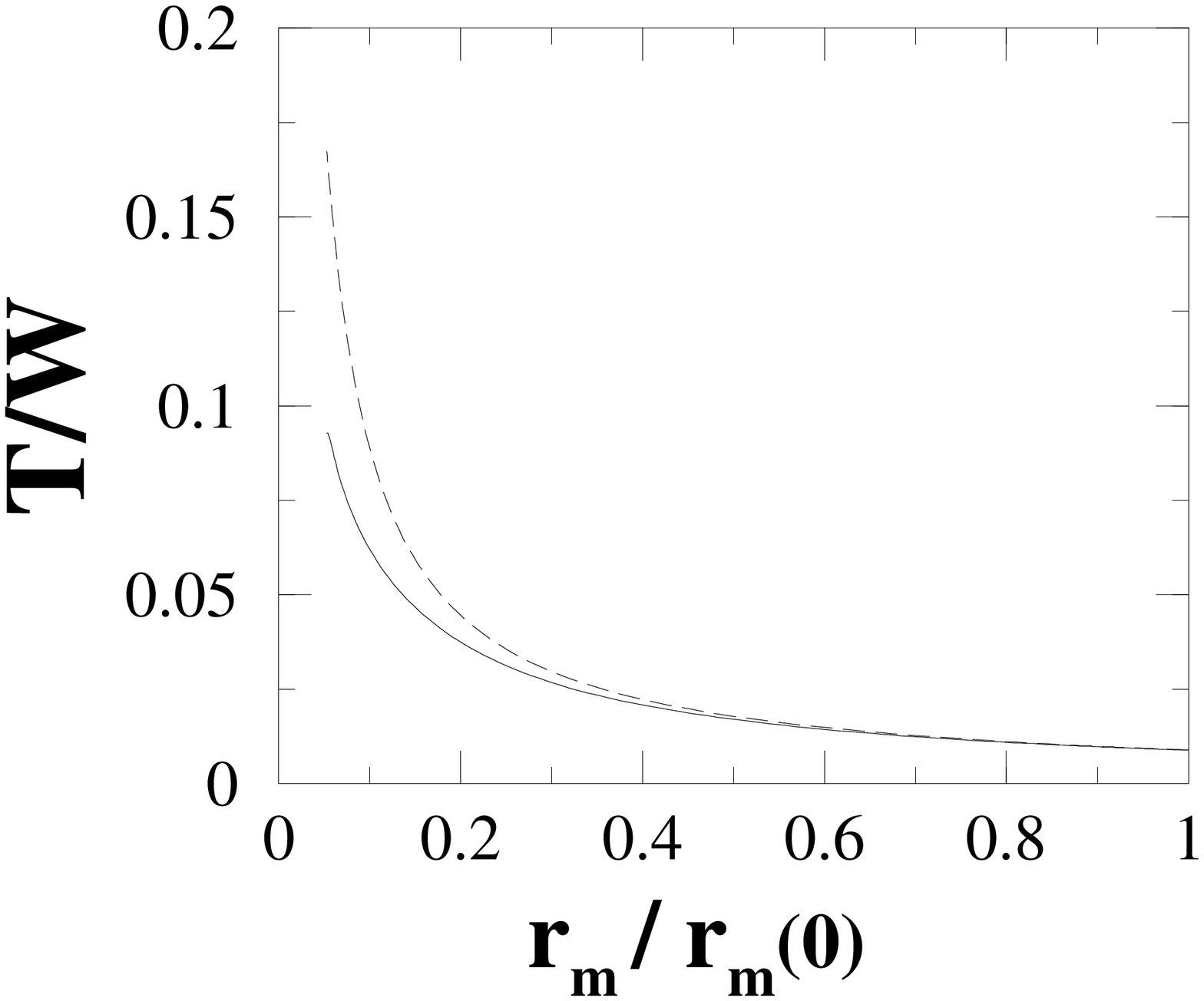}
\end{minipage}
\end{center}
\figcaption[f13.eps]{
Growth of $T/W$ during collapse.
The solid line shows the results of our simulation 
(the same $T/W$ as in Fig. \ref{fig:RCbar}). The
dash line assumes $T/W \propto 1/r_{m}$ during the 
collapse; this scaling would hold if the collapse were spherical.
Note that $r_{m}(0)$ is the mean radius 
at $t=0$.
\label{fig:TWscale}}
\vskip 8pt
\end{minipage}

\newpage

\begin{minipage}[t]{\minitwocolumn}
during the
evolution, hence we conclude that  a nonaxisymmetric bar does not
grow prior to BH formation.   Even 
though $T/W$ may exceed $\approx
0.27$ in the final stages, there is not sufficient time for growth prior
to the appearance of a hole.  The reason for the absence of an
instability is that $T/W$ does not grow as rapidly as $r_{m}^{-1}$ 
during the late stage of collapse, as it would remain spherical (see 
Fig. \ref{fig:TWscale}).

\hspace{3mm}
We show the density profile in the equatorial plane in Figure
\ref{fig:RCxy} and in the meridional plane in Figure \ref{fig:RCxz}.  The
ability of our scale factor implementation to resolve the matter 
distribution even as it becomes increasingly compact during the
implosion  is evident from these snapshots.  We find no indication of
the formation of a circumstellar disk with significant mass by  the
termination of our simulation. In fact, the fraction of the rest mass 
outside a sphere of radius 
$r/M_{0}=7.0$ is $26$\% and outside the sphere of $r/M_{0}=28.0$ is $10$\%.  
Accordingly,
most of the mass is concentrated in the center and is collapsing inward
when we terminate our integration. Note that by
employing a density cutoff, we are not reliably resolving the very
outermost region. But we note that even with a cutoff, $M_{0}$ is
conserved to 96\% accuracy (Fig.
\ref{fig:qp57vvcons}).  We also note that the overall picture is not
affected by a change of the cutoff value or extension of the grid size. 
We thus conclude that the rotation cannot provide sufficient
centrifugal support in the bulk of the envelope to counter gravity and
form a significant disk.
\end{minipage}
\hspace{\columnsep}
\begin{minipage}[t]{\minitwocolumn}
\hspace{3mm}
Though our computation is terminated when the lapse drops below
$\alpha_{\rm c} \sim 0.3$, we can still infer the final fate of the
collapse from examination of the velocity profile of the star (Figs.
\ref{fig:RCvxy} and \ref{fig:RCvxz}).   The growth of an appreciable
inward radial component of the velocity field strongly suggests that
immediately after the time we terminate the integrations the bulk of
the matter will cross the event horizon of the nascent BH in a
dynamical timescale as  measured at the center of the star.

\hspace{3mm}
Though the newly formed BH acquires the bulk of the mass in a 
coherent implosion, it does not obtain all of the mass and angular
momentum. Lingering gaseous fragments in the outermost envelope
containing negligible mass but nonnegligible angular momentum are
not followed in our simulation, which focuses on the imploding
massive bulk of the star. These fragments may accrete on a longer
timescale (or even escape), but we cannot track their evolution with
our current calculation.

\hspace{3mm}
Figure \ref{fig:RCr} shows the angular momentum distribution during
the evolution.  During the collapse, the central layers begin rotating
faster than the surface layers and the configuration, uniformly 
rotating initially, acquires appreciable differential rotation.

\hspace{3mm}
If all of the initial mass-energy and angular momentum are
consumed by the final BH, it will be rapidly rotating with $a/M
\approx 1$ (see Table \ref{tbl:RCInitial}).  Although we cannot follow
the final formation and growth of the BH, our PN simula-
\end{minipage}

\begin{center}
\begin{minipage}{13cm}
\epsfxsize 13cm
\epsffile{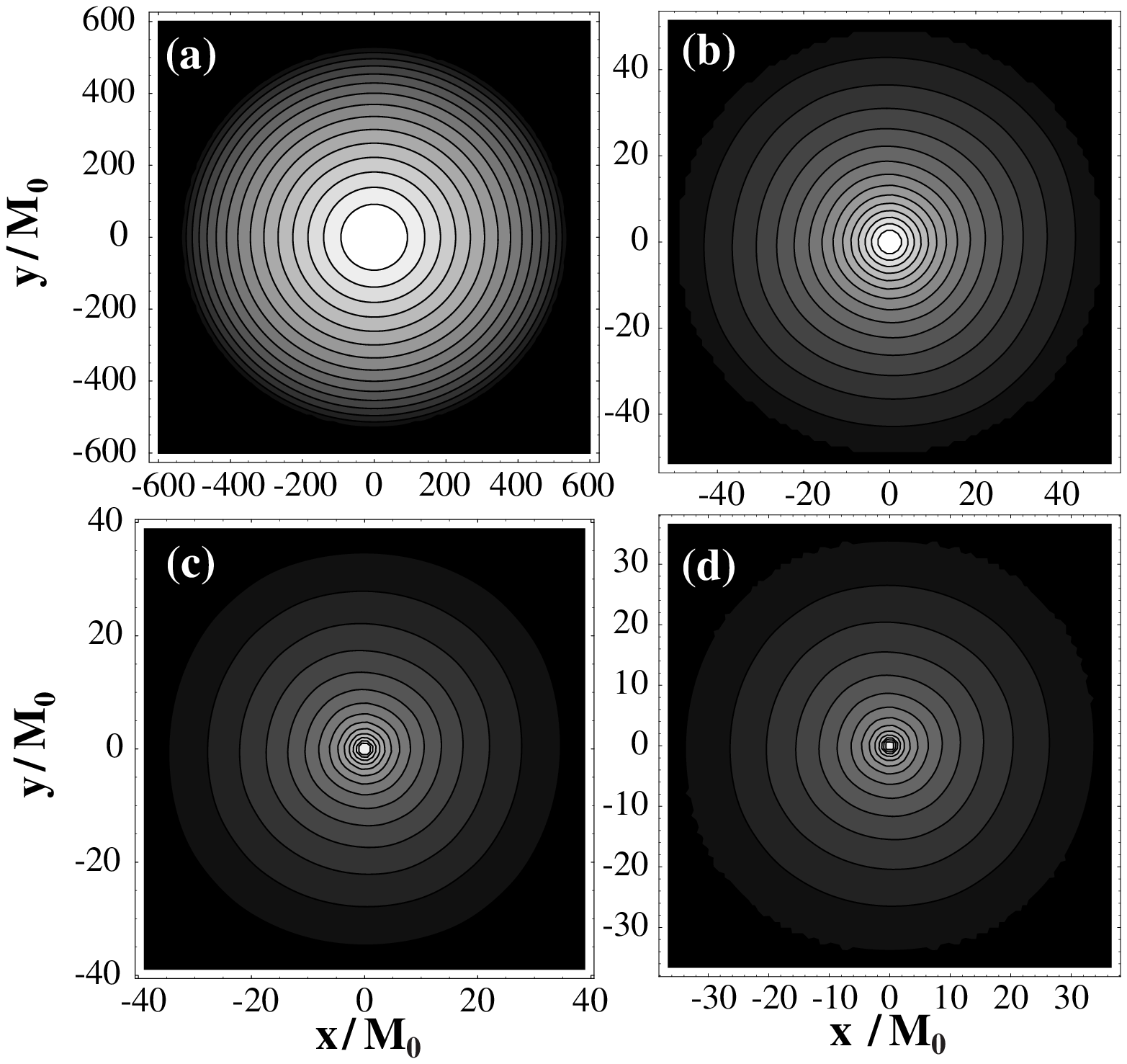}
\end{minipage}
\end{center}
\figcaption[f14.eps]{
Density contours $\rho_{*}$ in the equatorial plane at selected times
during rotating
SMS collapse.   Snapshots are plotted at ($t/t_{\rm D}$,
$\rho_{\rm c}^{*}$, $d$) = 
(a)  ($5.0628 \times 10^{-4}$, $8.254 \times 10^{-9}$, $10^{-7}$), 
(b) ($2.50259$, $1.225 \times 10^{-4}$, $10^{-5}$), 
(c) ($2.05360$, $8.328 \times 10^{-3}$, $5.585 \times 10^{-7}$), 
(d) ($2.50405$, $3.425 \times 10^{-2}$, $1.357 \times 10^{-7}$), respectively.  
The contour lines denote densities $\rho^{*}=\rho^{*}_{\rm c} \times
d^{(1-i/16)}$ ($i=1, \cdots , 15$).
\label{fig:RCxy}}
\vskip 12pt

\newpage

\begin{minipage}[t]{\minitwocolumn}
tions suggest that the final $a/M$ may be slightly lower, 

\begin{center}
\begin{minipage}{7.0cm}
\epsfxsize 6.5cm
\epsffile{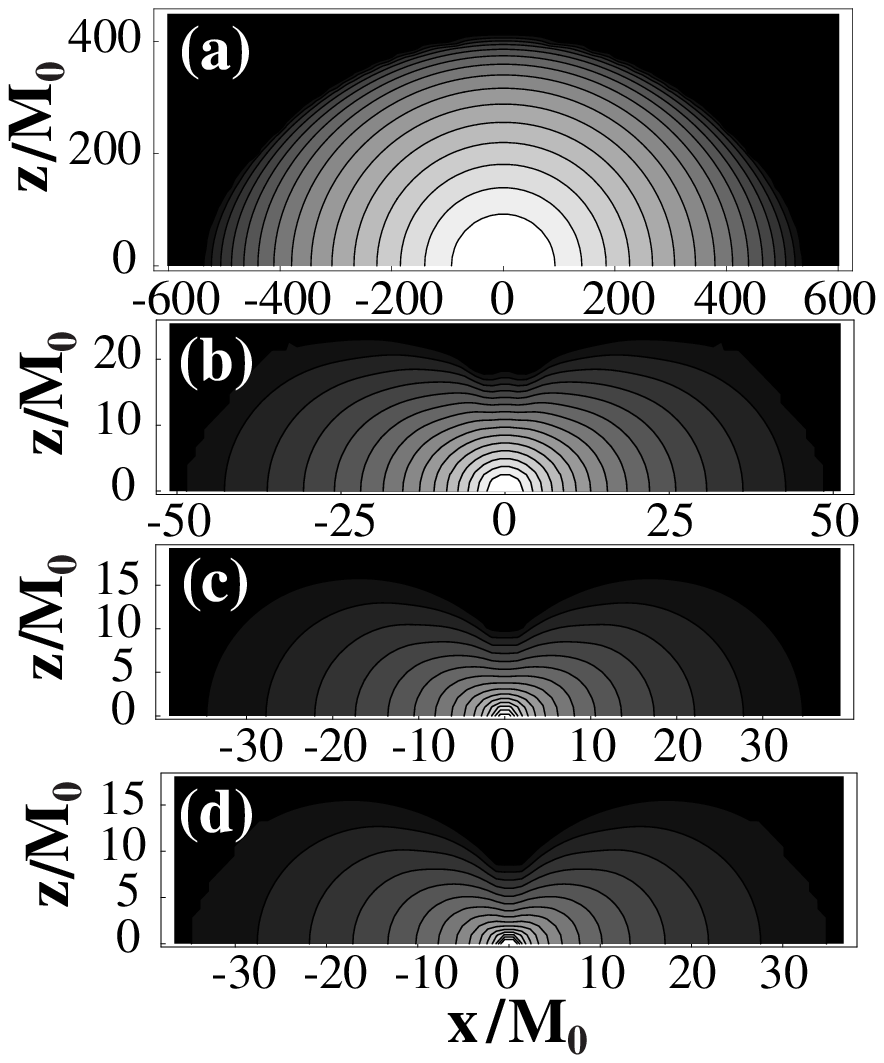}
\end{minipage}
\end{center}
\figcaption[f15.eps]{
Density contours $\rho_{*}$ in the meridional plane at selected times 
during rotating
SMS collapse.   The times, the central densities and
contour levels are the same as in Fig. \ref{fig:RCxy}.
\label{fig:RCxz}}
\vskip 12pt

\end{minipage}
\hspace{\columnsep}
\begin{minipage}[t]{\minitwocolumn}
due to the loss
of angular momentum carried by gas orbiting near the equator.

\section{Discussion}
\label{sec:Discussion}
\hspace{3mm}
We follow rotating SMS collapse from the onset of radial collapse at
$R_{p}/M_{0} \sim 411$ to the point where the PN approximation
breaks down ($R_{p}/M_{0} \sim 8$).  The challenge of covering this
large a dynamic  range is met by introducing a scale factor and a
``comoving'' coordinate system which takes advantage of the
homologous nature of the initial collapse.

\hspace{3mm}
Collapse of a uniformly rotating, relativistically unstable SMS is
coherent and leads to the formation of a SMBH containing the bulk of
the mass of the progenitor star.  It is interesting to contrast our
results with those of \citet{LR}, who treat the isothermal
($\Gamma=1$) collapse of initially homogeneous, uniformly
rotating, low entropy clouds via smooth particle hydrodynamics
(SPH) simulations.  They find considerable fragmentation into dense
clumps,  and disk formation containing $\sim 5\%$ of the mass. They
conclude that a seed BH will form at the center and that it likely
will grow gradually by accretion.

\end{minipage}

\begin{center}
\begin{minipage}{13cm}
\epsfxsize 13cm
\epsffile{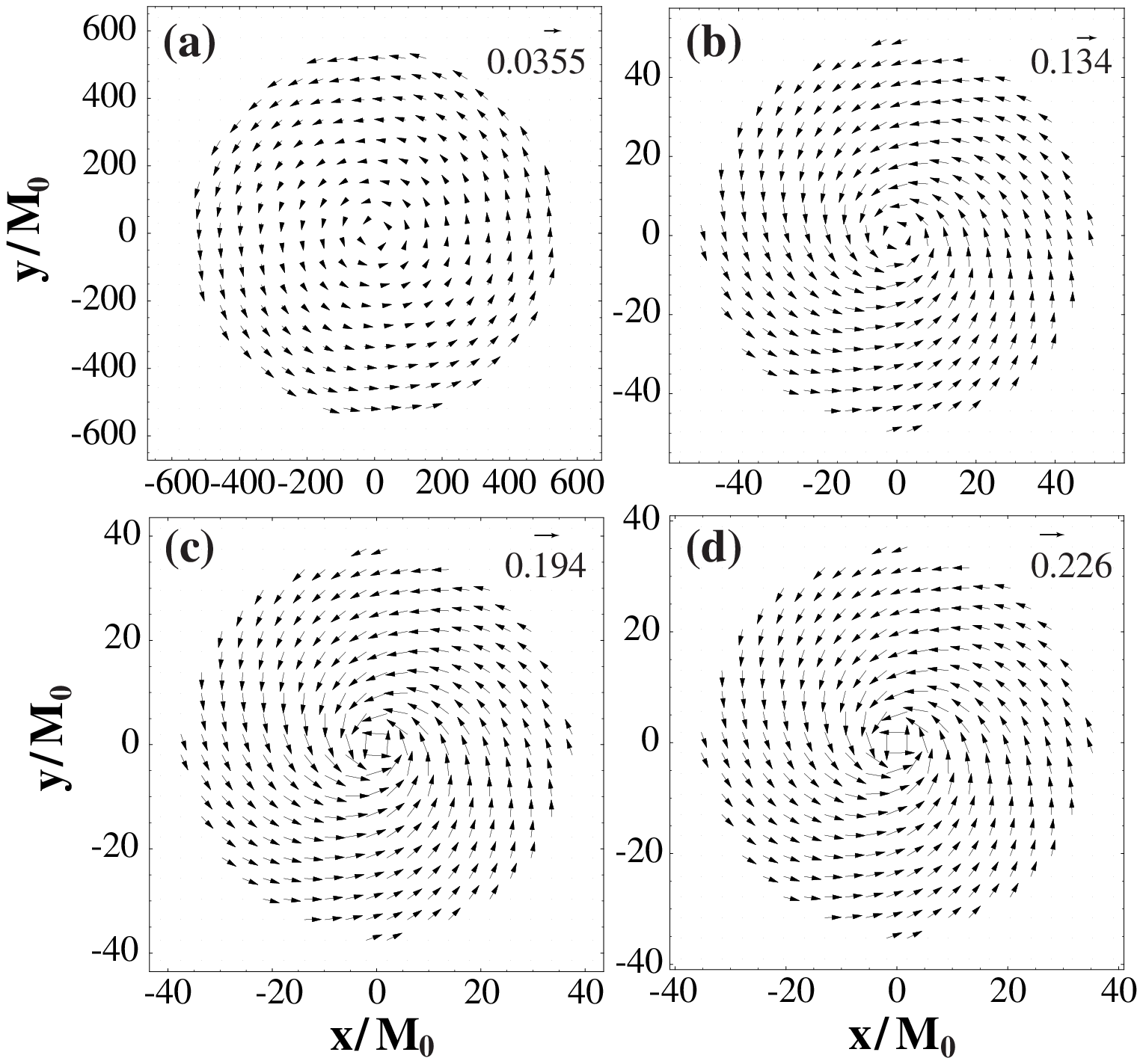}
\end{minipage}
\end{center}
\figcaption[f16.eps]{
Velocity field in the equatorial plane at selected times during 
rotating SMS collapse.   The
time for each snapshot is the same as in Fig.
\ref{fig:RCxy}. Note that the velocity field is drawn in 
physical coordinates and the velocity arrows are normalized as
indicated in the upper right hand corner of each snapshot.
\label{fig:RCvxy}}
\vskip 12pt

\newpage

\begin{minipage}[t]{\minitwocolumn}
\hspace{3mm}
We find no evidence of bars prior to BH formation, so that the
collapse is largely axisymmetric. As a result, little angular
momentum can be radiated away by gravitational waves.  

\hspace{3mm}
From the coherent, axisymmetric  nature of the implosion we conclude
that the collapse of a SMS, rotating uniformly at the onset of collapse,
is a promising source of  gravitational wave bursts.  
We can estimate
the strength and the frequency of the wave burst emitted from this

\begin{center}
\begin{minipage}{7.0cm}
\epsfxsize 6.5cm
\epsffile{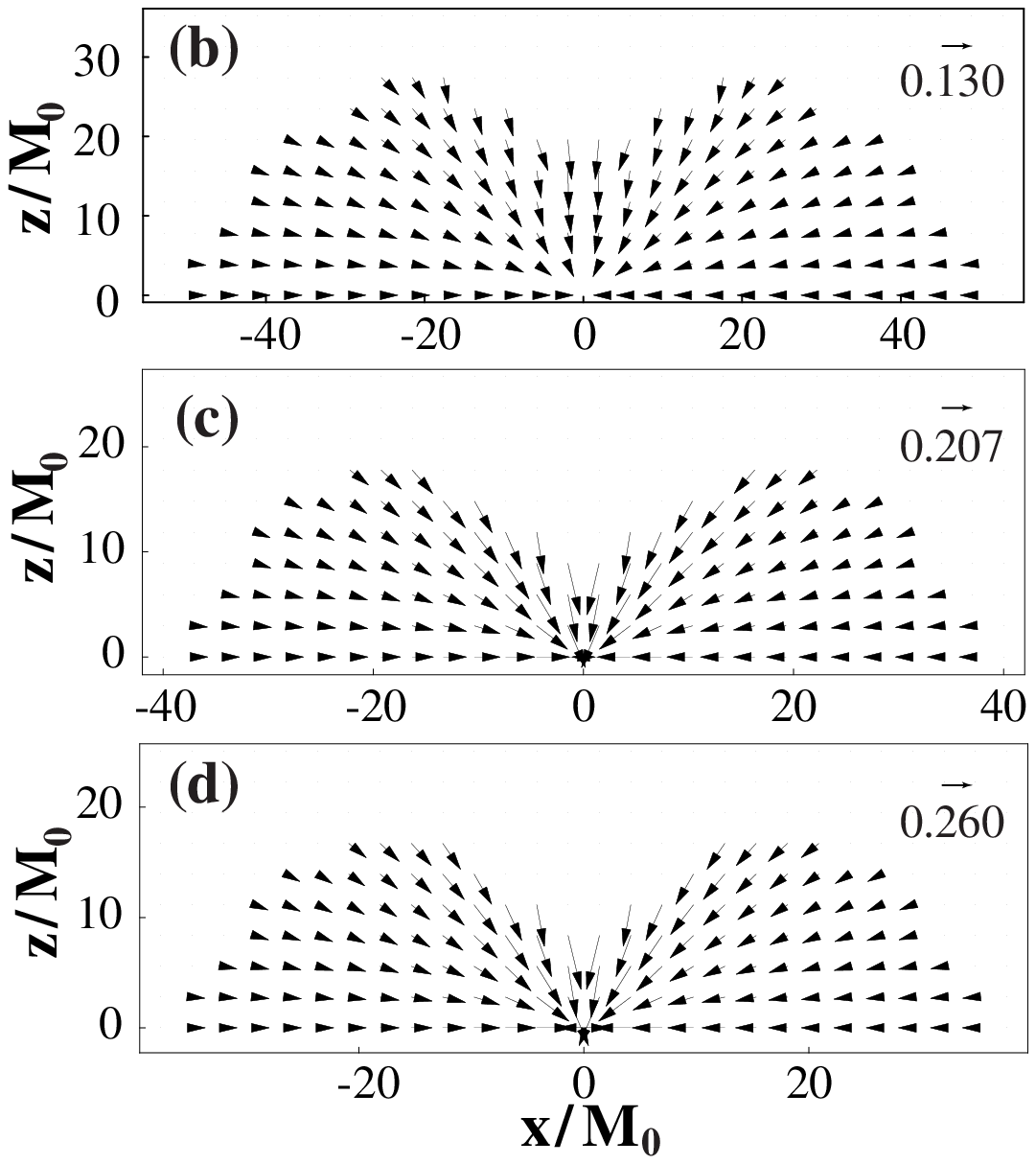}
\end{minipage}
\end{center}
\figcaption[f17.eps]{
Velocity fields in the meridional plane at selected times during
rotating SMS collapse.  
The time for each snapshot is the same to Fig.
\ref{fig:RCxy}. We omit Fig. \ref{fig:RCvxz} (a) because
there is no velocity in the meridional plane at $t=0$.
\label{fig:RCvxz}}
\vskip 12pt
\end{minipage}
\hspace{\columnsep}
\begin{minipage}[t]{\minitwocolumn}
rotating collapse. The characteristic burst frequency is given
by  the
dynamical timescale of the star at the time of BH formation,
\begin{eqnarray}
f_{\rm burst} &\sim& \frac{1}{2 \pi t_{\rm dyn}} 
\sim \frac{1}{2\pi} \left( \frac{M}{R^{3}} \right)^{1/2}
\nonumber \\
&= &
3 \times 10^{-3} 
\left( \frac{10^{6} M_{\odot}}{M} \right) 
\left( \frac{5M}{R} \right)^{3/2}
[{\rm Hz}].
\end{eqnarray}
The wave amplitude can be estimated by employing the Newtonian quadrupole
formula according to
\begin{eqnarray}
h_{\rm burst} &\sim& \frac{\ddot{Q}}{d} \sim
\frac{MR^{2}f_{\rm burst}^{2}}{d} 
\sim \frac{M}{d} \frac{1}{4\pi^{2}} \frac{M}{R}
\nonumber \\
&=& 2 \times 10^{-19} \left( \frac{M}{10^{6} M_{\odot}} \right)
\left( \frac{1 {\rm G pc}}{d} \right)
\left( \frac{5M}{R} \right)
,
\end{eqnarray}
where $Q$ is the quadrupole moment of the star and $d$ is the distance
from the observer. We set $R/M=5$, a characteristic mean radius during
BH formation.  Since the main targets of LISA are gravitational
radiation sources between $10^{-4}$ and $10^{-1}$ Hz, it is possible
that LISA can search for the burst waves accompanying rotating SMS
collapse and formation of a SMBH.

\hspace{3mm}
In the absence of bar formation, SMS collapse will not produce
quasi-periodic waves prior to SMBH formation.  However, such
waves will be generated by the nascent BH via quasi-normal mode
ringing.  The characteristic frequency
$f_{\rm QNM}$ and strength $h_{\rm QNM}$ of this radiation in 
rotating star collapse are \citep*{Thorne87,SSU}
\end{minipage}

\begin{center}
\begin{minipage}{13cm}
\epsfxsize 13cm
\epsffile{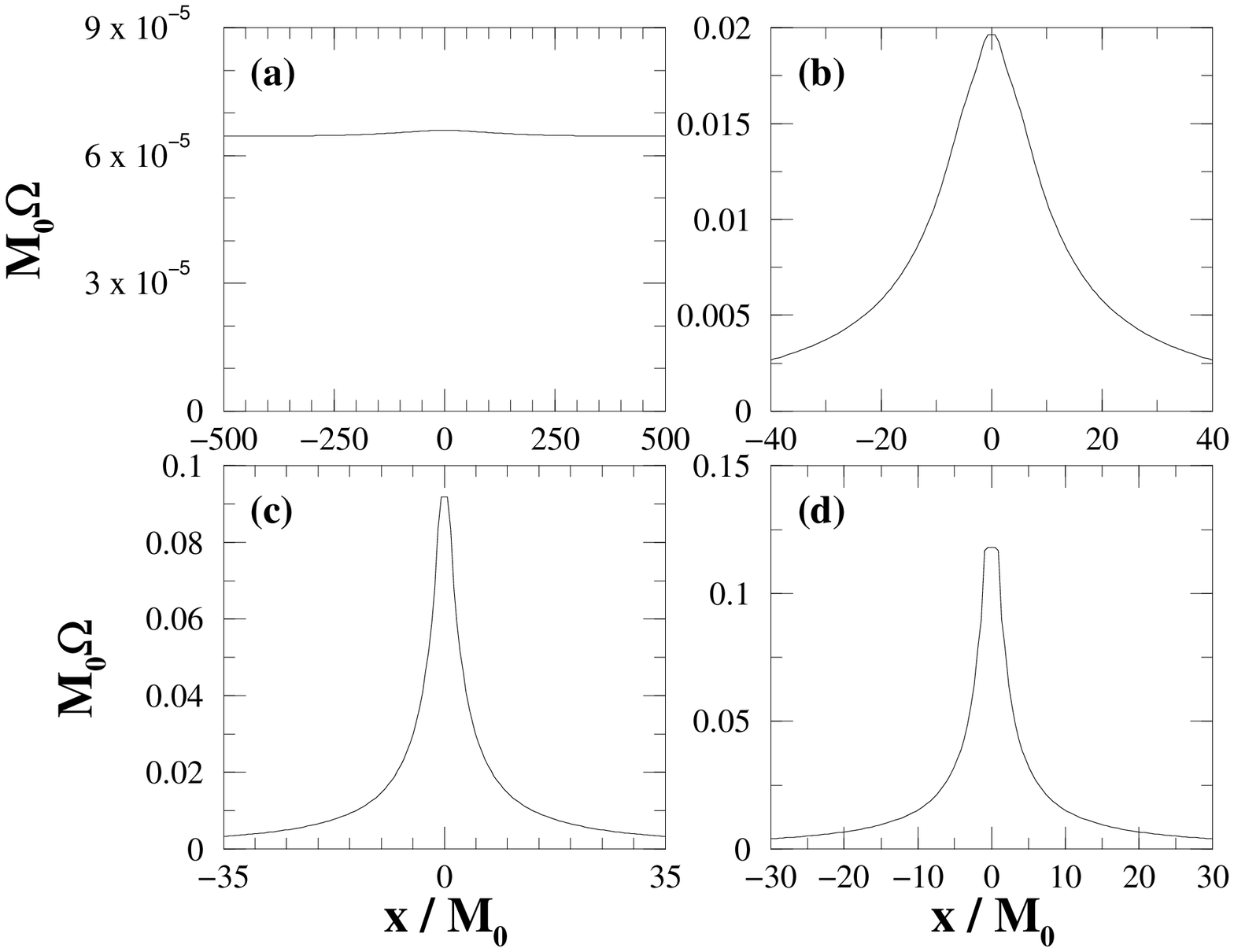}
\end{minipage}
\end{center}
\figcaption[f18.eps]{
Angular momentum distribution along the $x$-axis at selected times 
during rotating SMS collapse. The time for each snapshot is the same
as Fig. \ref{fig:RCxy}.
\label{fig:RCr}}
\vskip 12pt

\begin{minipage}[t]{\minitwocolumn}
\begin{eqnarray}
f_{\rm QNM} &\sim&
2 \times 10^{-2} \left( \frac{10^{6} M_{\odot}}{M} \right) {\rm [Hz]}
, \\
h_{\rm QNM} &\sim&
6 \times 10^{-19} \left( \frac{\Delta E_{\rm GW}/M}{10^{-4}} \right)^{1/2}
\left( \frac{2 \times 10^{-2} {\rm [Hz]}}{f_{\rm QNM}} \right)^{1/2}
\nonumber \\
&& \times
\left( \frac{M}{10^{6} M_{\odot}} \right)^{1/2}
\left( \frac{1 {\rm Gpc}}{d} \right)
,
\end{eqnarray}
where $\Delta E_{\rm GW}$ is total radiated energy.  Here we use the
$l=m=2$ quasi-normal mode frequency $\omega_{\rm QNM}$ for a Kerr
BH with $a/M=0.9$, for which $M \omega_{\rm QNM}=0.7$
\citep{Leaver}.  The efficiency of the radiated energy could be less than
$\Delta E_{\rm GW} / M \lesssim 7 \times 10^{-4}$ for a rotating
collapse \citep{SP}.  From the above estimate, gravitational waves
from the vibration of a newly formed SMBH reside within the
sensitivity limits of LISA.

\hspace{3mm}
Since we assume that the equilibrium SMS is uniformly rotating 
initially, it cannot support a large amount of angular momentum
without exceeding the mass-shedding limit.  However, when internal
magnetic fields and viscosity are weak, the equilibrium star may
rotate differentially and support a considerably larger angular
momentum (and higher $T/W$) without exceeding the mass-shedding
limit.  For such a case, \citet{NS} propose an alternative scenario for
the quasi-static cooling and contraction of an  equilibrium SMS which
inevitably leads to bar formation prior to catastrophic collapse. Such a
scenario will almost certainly generate long wavelength quasi-periodic
wave emission, even in the absence of SMBH formation.

\hspace{3mm}
Our (3+1) hybrid PN calculations offer a first glance at SMS collapse.
An improved description will require several refinements to our
computational scheme.  
First, it will be necessary to employ a fully
relativistic treatment 
\end{minipage}
\hspace{\columnsep}
\begin{minipage}[t]{\minitwocolumn}
of Einstein's field equations to explore the final
dynamical phase of collapse  once a BH has formed.  
In fact, we cannot
reliably use PN gravity once the radius decreases much below
$R_{p}/M \lesssim 20$, since the central fields are becoming strong. One possibility would be
to use our output when the central fields are beginning to get strong 
($\alpha_c \lesssim 0.5$) as initial data for a fully relativistic
(3+1) code.  This would allow us to take advantage of the faster, more
robust PN code to handle the Newtonian and PN implosion regime and
switch over to a fully general relativistic code only for the final
strong-field phase of collapse.

\hspace{3mm}
Secondly, our computation would benefit significantly from nested grid 
method \citep[see, e.g.,][]{Ruffert} to handle the large dynamic range 
characterizing SMS collapse.  We have successfully exploited approximate 
homology by introducing a scale factor to set up
an approximate comoving coordinate system. But homology breaks
down at the center toward the end of the collapse. Moreover it is
difficult to follow the small amount of matter in the equatorial plane
that is supported by centrifugal forces while the bulk of the matter
collapses.  Nested grid method may provide one means of 
concentrating computational
resources on the central region of the star while simultaneously
resolving  the low density outermost regions.

\acknowledgements
\hspace{3mm}
We thank Matt Duez and H.-Thomas Janka for their critical reading of our 
manuscript.  The computations reported here were performed under vector computer
NEC-SX5 at the Yukawa Institute for Theoretical Physics, Kyoto
University.   This work was supported in part by NSF Grants
PHY-0090310 and 99-02833 and NASA Grants NAG5-8418  and
NAG5-10781 at the University of Illinois at Urbana-Champaign, 
and a Japanese Monbu-Kagakusho Grant, No. 13740143.

\end{minipage}


\end{document}